\documentclass[prb,twocolumn,showpacs,showkeys,floatfix]{revtex4}

\usepackage{graphicx}
\usepackage{dcolumn}
\usepackage{bm}
\usepackage{subfigure}
\usepackage{array}
\usepackage{float}
\usepackage{enumerate}
\usepackage{multirow}
\usepackage{amsmath}
\usepackage{amssymb}
\usepackage{epstopdf}

\newcommand{\fig}[1]{FIG.~\ref{#1}}
\newcommand{\tab}[1]{TABLE~\ref{#1}}

\begin{document}

\title{Thermodynamic properties of binary HCP solution phases from special quasirandom structures}

\author{Dongwon Shin}\email{dus136@psu.edu}
\author{Raymundo Arr\'{o}yave}
\author{Zi-Kui Liu}
\affiliation{Department of Materials Science and Engineering, The Pennsylvania
State University, University Park, Pennsylvania 16802}
\author{Axel Van de Walle}
\affiliation{Engineering and Applied Science Division, California Institute of
Technology, Pasadena, California 91125, USA}

\date{\today}

\begin{abstract}
Three different special quasirandom structures (SQS) of the substitutional hcp
$A_{1-x}B_x$ binary random solutions ($x=0.25$, $0.5$, and $0.75$) are
presented. These structures are able to mimic the most important pair and
multi-site correlation functions corresponding to perfectly random hcp
solutions at those compositions. Due to the relatively small size of the
generated structures, they can be used to calculate the properties of random
hcp alloys via first-principles methods. The structures are relaxed in order to
find their lowest energy configurations at each composition. In some cases, it
was found that full relaxation resulted in complete loss of their parental
symmetry as hcp so geometry optimizations in which no local relaxations are
allowed were also performed. In general, the first-principles results for the
seven binary systems (Cd-Mg, Mg-Zr, Al-Mg, Mo-Ru, Hf-Ti, Hf-Zr, and Ti-Zr) show
good agreement with both formation enthalpy and lattice parameters measurements
from experiments. It is concluded that the SQS's presented in this work can be
widely used to study the behavior of random hcp solutions.
\end{abstract}



\maketitle

\section{Introduction}
\label{sec:intro}

Thermodynamic modeling using the CALPHAD (CALculation of PHAse Diagrams)
method~\cite{1970Kau,1998Sau} attempts to describe the Gibbs energy of a system
through empirical models whose parameters are fitted using experimental
information. These descriptions allow the extrapolation of a system's
thermodynamic properties to regions in the composition-temperature space that
have not/cannot be accessed through experiments. These empirical models,
however, are as good as the data used to fit them and are therefore limited by
the availability of accurate experimental data. This limitation can be overcome
by using theoretical calculations based on first-principles methods, which are
capable of predicting the physical properties of phases with no experimental
input~\cite{2002Wal}. Unfortunately, despite their predictive nature, these
methods are not yet able to calculate the thermochemistry of
materials---especially multi-component, multi-phase systems---with the
precision required in industry.

A natural way of improving the predictive capabilities of empirical models
while maintaining their applicability to practical problems is by combining
first-principles and CALPHAD techniques. Thanks to efficient schemes for
implementing Density Functional Theory (DFT)\cite{Koh65}, the almost-routine
use of first-principles results within the CALPHAD methodology has become a
reality. In this hybrid approach, the energetics obtained through electronic
structure calculations are used as input data within the CALPHAD formalism to
obtain the parameters that describe the Gibbs energy of the
system\cite{2002Wol}.

The first-principles electronic structure calculations of perfectly ordered
periodic structures are relatively straightforward since they usually rely on
the use of periodic boundary conditions. Problems arise, however, when
attempting to use these methods to study the thermochemical properties of
random solid solutions since an approximation must be made in order to simulate
a random atomic configuration through a periodic structure. The usual
approaches that have been used in the past can be summarized as follows:

\begin{enumerate}[i)]
\item
{The most direct approach is the supercell method. In this case, the sites of
the supercell can be randomly occupied by either A or B atoms to yield the
desired $A_{1-x}B_{x}$ composition. In order to reproduce the statistics
corresponding to a random alloy, such supercells must necessarily be very
large. This approach is, therefore, computationally prohibitive when the size
of the  supercell is on the order of hundreds of atoms.}

\item
{Another technique, the Coherent Potential Approximation (CPA)\cite{1967Sov}
method, is a single-site approximation that models the random alloy as an
ordered lattice of effective atoms. These are constructed from the criterion
that the average scattering of electrons off the alloy components should
vanish\cite{2005Kis}. In this method, local relaxations are not considered
explicitly and the effects of alloying on the distribution of local
environments cannot be taken into account. Local relaxations have been shown to
significantly affect the properties of random solutions~\cite{1991Lu},
especially when the constituent atoms vary greatly in size and, therefore,
their omission constitutes a major drawback. Although the local relaxation
energy can be taken into account\cite{2005Kis}, these corrections rely on
cluster expansions of the relaxation energy of ordered structures and the
distribution of local environments is not explicitly considered. Additionally,
such corrections are system specific.}

\item
{A third option is to apply the Cluster Expansion approach\cite{1993San}. In
this case, a generalized Ising model is used and the spin variables can be
related to the occupation of either atom A or B in the parent lattice. In order
to obtain an expression for the configurational energy of the solid phase, the
energies of multiple configurations (typically in the order of a few dozens)
based on the parent lattice must be calculated to obtain the parameters that
describe the energy of any given $A_{1-x}B_{x}$ composition. This approach
typically relies on the calculation of the energies of a few dozen ordered
structures.}
\end{enumerate}

In the techniques outlined above, there are serious limitations in terms of
either the computing power required (supercells, cluster expansion) or the
ability to accurately represent the local environments of random solutions
(CPA). Ideally, one would like to be able to accurately calculate the
thermodynamic and physical properties of a random solution with as small a
supercell as possible so that accurate first-principles methods can be applied.
This has become possible thanks to the development of Special Quasirandom
Structures (SQS).

The concept of SQS was first developed by \citet{1990Zun} to mimic random
solutions without generating a large supercell or using many configurations.
The basic idea consists of creating a small ---4-48 atoms--- periodic structure
with the target composition that best satisfies the pair and multi-site
correlation functions corresponding to a random alloy, up to a certain
coordination shell. Upon relaxation, the atoms in the structure are displaced
away from their equilibrium positions, creating a distribution of local
environments that can be considered to be representative of a random solution,
at least up to the first few coordination shells.

Provided the interatomic electronic interactions in a given system are
relatively short-range, the first-principles calculations of the properties of
these designed supercells can be expected to yield sensible results, especially
when calculating properties that are mostly dependent on the local atomic
arrangements, such as enthalpy of mixing, charge transfer, local relaxations,
and so forth. It is important to stress that the approach fails whenever a
property depends on long-range interactions.

The SQS's for fcc-based alloys and bcc alloys have been generated by
\citet{1990Wei} and \citet{2004Jia}, respectively. However, to the best
knowledge of these authors, there has been no investigation on the application
of the SQS approach to the study of hcp substitutional random solutions. In the
present work, we propose two SQS's capable of mimicking hcp random alloys at
25, 50 and 75 at.\%. The paper is organized as follows:

The proposed SQS are characterized in terms of their ability to reproduce the
pair and multi-site correlation functions of a truly random hcp solution.
Subsequently, the structures are tested in terms of their ability to reproduce,
via first-principles calculations, the properties of certain selected stable or
metastable binary hcp solutions, namely Cd-Mg, Mg-Zr, Al-Mg, Mo-Ru, Hf-Ti,
Hf-Zr and Ti-Zr. To further analyze the relaxation behavior of the structures,
the distribution of first nearest bond lengths as well as the radial
distribution for the first few coordination shells is presented. Finally, for
each of the selected binaries, the calculated and available experimental
lattice parameters and enthalpy of mixing are compared. Results from other
techniques are also presented where available in order to further corroborate
the present calculations.

\section{Generation of special quasirandom structures}%%
\label{sec:gensqs}

In order to characterize the statistics of a given atomic arrangement, one can
use its correlation function\cite{1991Ind}. Within the context of lattice
algebra, we can assign a "spin value", $\sigma=\pm 1$, to each of the sites of
the configuration, depending on whether the site is occupied by A or B-type
atoms. Furthermore, all the sites can be grouped in figures,
$f\left(k,m\right)$, of $k$ vertices, where $k$=1,2,3,$\cdots$ responds to a
shape, point, pair, and triplet,$\dots$ respectively, spanning a maximum
distance of $m$, where $m=1,2,3,\cdots$ is the first, second, and third-nearest
neighbors, and so forth. The correlation functions, $\overline{\Pi}_{k,m}$, are
the averages of the products of site occupations ($\pm 1$ for binary alloys and
$\pm 1$, 0 for ternary alloys) of figure \emph{k} at a distance \emph{m} and
are useful in describing the atomic distribution. The optimum SQS for a given
composition is the one that best satisfies the condition:

\begin{equation}
\left( \overline{\Pi}_{k,m} \right)_{SQS} \cong \langle \overline{\Pi}_{k,m}
\rangle_R
\end{equation}

\noindent where $\langle \overline{\Pi}_{k,m} \rangle_R$ is the correlation
function of a random alloy, which is simply by $(2x-1)^k$ in the $A_{1-x}B_x$
substitutional binary alloy, where $x$ is the composition. We considered SQS's
of two different compositions, i.e. $x=0.5$ and $0.75$.

Unlike cubic structures, the order of a given configuration in the hcp lattices
relative to a given lattice site may be altered with the variation of c/a
ratio. However, these new arrangements will not cause any change in the
correlation functions, since one can thus use \emph{any} c/a ratio to generate
the hcp SQS's. As a matter of simplicity, the ideal c/a ratio was considered in
order to generate SQS's.

\begin{figure*}[hbt]
\centering %
    \subfigure[~SQS-16 for $x$=0.5]{
        \label{fig:5050}
        \includegraphics[width=3.0in]{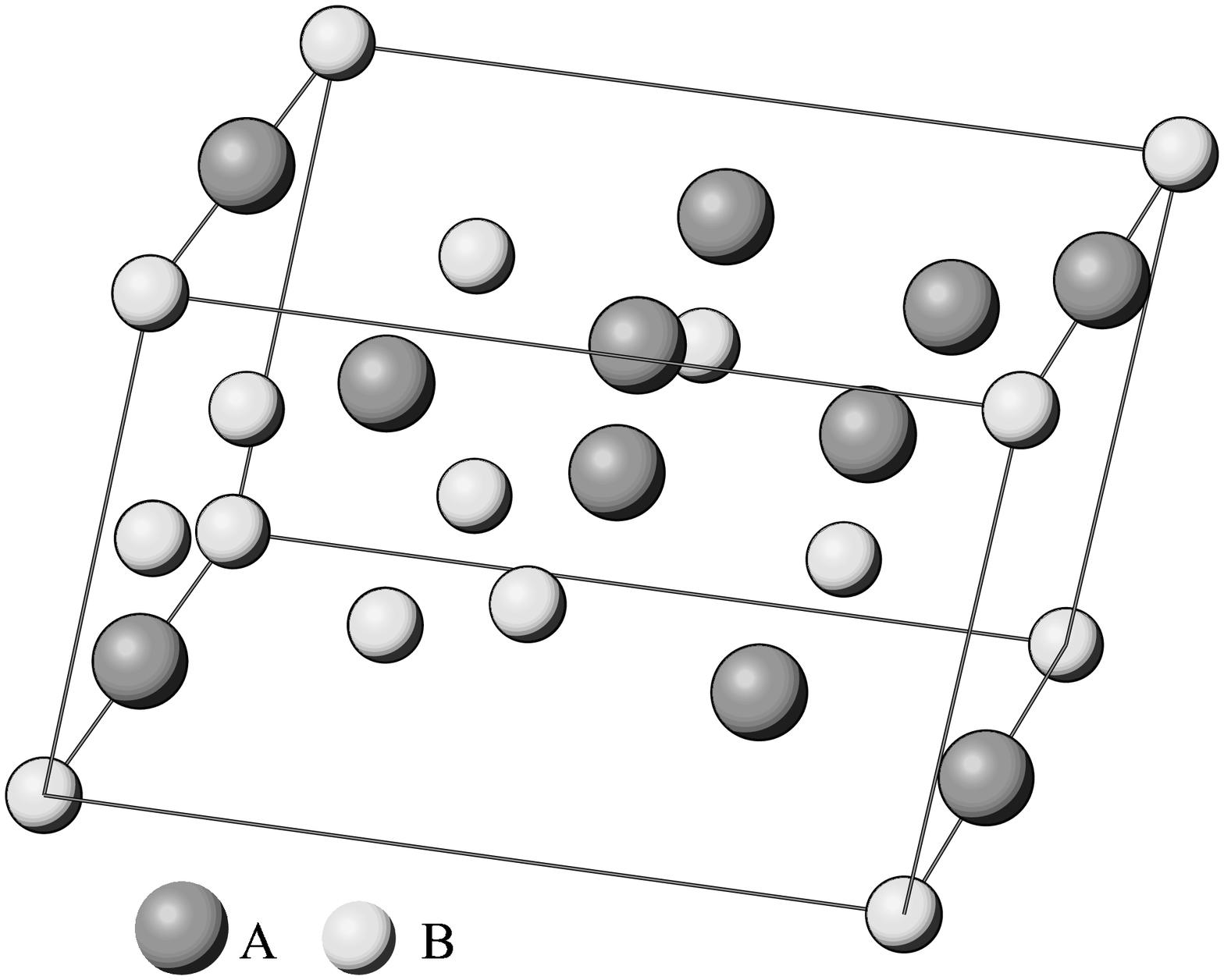}
        }
    \subfigure[~SQS-16 for $x$=0.75]{
        \label{fig:2575}
        \includegraphics[width=3.0in]{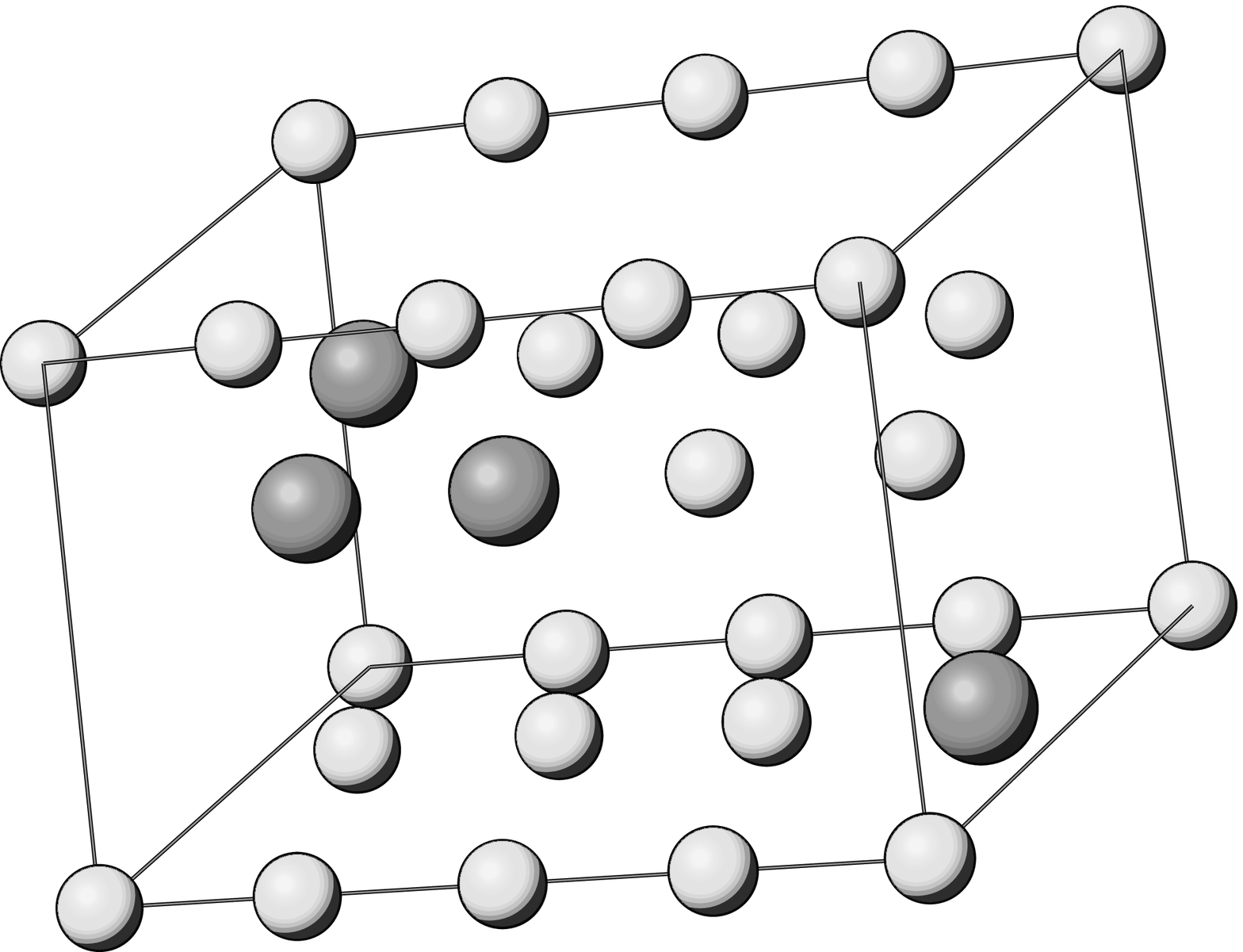}
    }
\caption{%
Crystal structures of the $A_{1-x}B_x$ binary hcp SQS-16 structures in
their ideal, unrelaxed forms. All the atoms are at the ideal hcp sites, even
though both structures have the space group, P1.
}%
\label{fig:sqs}
\end{figure*}

In the present work, we used the Alloy Theoretic Automation Toolkit
(ATAT)\cite{2002Wal} to generate special quasirandom structures for the hcp
structure of 8 and 16 sites. The schematic diagrams of the created special
quasirandom structure with 16 atoms are shown in \fig{fig:sqs} and the
corresponding lattice vectors and atomic positions are listed in
\tab{tbl:lattice}.

The correlation functions of the generated 8 and 16-atom SQS's were
investigated to verify that they satisfied at least the short-range statistics
of an hcp random solution. As is shown in \tab{tbl:Correlation}, the 16-atom
structures satisfy the pair correlation functions of random alloys up to the
fifth and third nearest neighbor for the 50 at.\% and the 75 at.\%
compositions, respectively. On the other hand, \tab{tbl:Correlation} shows that
the SQS-8 for 75 at.\% could not satisfy the random correlation function even
for the first-nearest neighbor pair. Thus, SQS's with 16 atoms are capable of
mimicking a random hcp configuration beyond the first coordination shell.

It is important to note that in \tab{tbl:Correlation}, and contrary to what is
observed in the SQS for cubic structures, some figures have more than one
crystallographically inequivalent figure at the same distance. For example, in
the case of hcp lattices with the ideal c/a ratio, two pairs may have the same
interatomic distance and yet be crystallographically inequivalent. In this
case, despite the fact that the two pairs (0,0,0) and ($a$,0,0); (0,0,0) and
($\tfrac13$,$\tfrac23$,$\tfrac12$), have the same inter-atomic distance, $a$,
they do not share the same symmetry operations. This degeneracy is broken when
the c/a ratio deviates from its ideal value.

For the sake of efficiency, the initial lattice parameters of the SQS's were
determined from Vegard's law. By doing so, the c/a ratio was no longer ideal.
Afterwards, we checked the correlation functions of the \emph{new} structures
and found that they remained the same as long as the corresponding figures were
identical.

\setlength{\extrarowheight}{3pt}
\begin{table}[htb]
\centering %
\fontsize{9}{9pt}\selectfont %%
\caption{\label{tbl:lattice}%
Structural descriptions of the SQS-$N$ structures for the binary hcp solid
solution. Lattice vectors and atomic positions are given in fractional
coordinates of hcp lattice. Atomic positions are given for the ideal, unrelaxed
hcp sites.}
\begin{ruledtabular}
\begin{tabular}{ccc}
 & $x=0.5$ & $x=0.75$\\
\hline
\multirow{21}{*}{SQS-16} & Lattice vectors & Lattice vectors\\
 & $\left(\begin{array}{rrr}
   0 & -1 & -1 \\
  -2 & -2 &  0 \\
  -2 &  1 & -1 \\
\end{array}\right)$
 & $\left(\begin{array}{rrr}
   1 &  1 & 1 \\
  -1 &  0 & 1 \\
   0 & -4 & 0 \\
\end{array}\right)$
 \\
 & Atomic positions & Atomic positions\\
 &
$\begin{array}{rrrc}
-2\tfrac13 & -1\tfrac23 & -1\tfrac12 & A \\
-1         & -1         & -1         & A \\
-2         &  0         & -1         & A \\
-1\tfrac13 & -\tfrac23  & -1\tfrac12 & A \\
-3         & -2         & -1         & A \\
-2\tfrac13 & -\tfrac23  & -1\tfrac12 & A \\
-4         & -2         & -2         & A \\
-3\tfrac13 & -1\tfrac23 & -1\tfrac12 & A \\
-2         & -2         & -1         & B \\
-1\tfrac13 & -1\tfrac23 & -\tfrac12  & B \\
-3         & -1         & -1         & B \\
-2         & -1         & -1         & B \\
-1\tfrac13 & -\tfrac23  & -\tfrac12  & B \\
-\tfrac13  & -\tfrac23  & -\tfrac12  & B \\
-2\tfrac13 & -1\tfrac23 & -\tfrac12  & B \\
-3         & -1         & -2         & B
\end{array}$
 &
$\begin{array}{rrrc}
-\tfrac13  & -2\tfrac23 & 1\tfrac12  & A \\
-\tfrac13  & -1\tfrac23 & 1\tfrac12  & A \\
0          & -3         & 2          & A \\
0          & -3         & 1          & A \\
0          & -2         & 2          & B \\
0          & -1         & 2          & B \\
0          & 0          & 2          & B \\
-\tfrac13  & -\tfrac23  & 1\tfrac12  & B \\
-\tfrac13  & \tfrac13   & 1\tfrac12  & B \\
-\tfrac13  & -3\tfrac23 & \tfrac12   & B \\
0          & -2         & 1          & B \\
-\tfrac13  & -2\tfrac23 & \tfrac12   & B \\
0          & -1         & 1          & B \\
-\tfrac13  & -1\tfrac23 & \tfrac12   & B \\
0          & 0          & 1          & B \\
-\tfrac13  & -\tfrac23  & \tfrac12   & B
\end{array}$\\
\\\hline\\
 \multirow{11}{*}{SQS-8} &Lattice vectors & Lattice vectors\\

 & $\left(\begin{array}{rrr}
  -1 &  1 & 1 \\
   1 & -1 & 1 \\
   1 &  1 & 0 \\
\end{array}\right)$
& $\left(\begin{array}{rrr}
%   1 &  1 & -1 \\
%   0 & -1 & -1 \\
%  -2 &  2 & 0 \\
   0 &  1 & 1 \\
  -1 &  0 & 1 \\
   2 & -2 & 0 \\
\end{array}\right)$\\
 & Atomic positions & Atomic positions\\

 &
 $\begin{array}{rrrc}
0        & 1         & 1         & A \\%
\tfrac23 & 1\tfrac13 & \tfrac12  & A \\%
\tfrac23 & 1\tfrac13 & 1\tfrac12 & A \\%
1        & 1         & 2         & A \\%
\tfrac23 & \tfrac13  & \tfrac12  & B \\%
\tfrac23 & \tfrac13  & 1\tfrac12 & B \\%
1        & 0         & 1         & B \\%
1        & 1         & 1         & B
 \end{array}$
 &
 $\begin{array}{rrrc}
 \tfrac23  & -\tfrac23  & \tfrac12  & A \\%
 1         & -1         & 1         & A \\%
-\tfrac13  & \tfrac13   & 1\tfrac12 & B \\%
 0         & 0          & 1         & B \\%
 0         & 0          & 2         & B \\%
 \tfrac23  & -\tfrac23  & 1\tfrac12 & B \\%
 1         & -1         & 2         & B \\%
 1\tfrac23 & -1\tfrac23 & \tfrac12  & B
 \end{array}$\\
\end{tabular}
\end{ruledtabular}
\end{table}

\begin{table*}[bth]
\fontsize{8}{8pt}\selectfont %%
\caption{\label{tbl:Correlation}Pair and multi-site correlation functions of
SQS-$N$ structures when the c/a ratio is ideal. The number in the square
bracket next to $\overline{\Pi}_{k,m}$ is the number of equivalent figures at
the same distance in the structure, the so-called degeneracy factor.}
\begin{ruledtabular}
\begin{tabular}{lccccccccc}
\multicolumn{1}{c}{} & \multicolumn{3}{c}{$x$=0.5} & \multicolumn{3}{c}{$x$=0.75}\\
 & Random & SQS-16 & SQS-8 & Random & SQS-16 & SQS-8\\
\hline
$\overline{\Pi}_{2,1}$[6] & 0 &  0 & 0 & 0.25 & 0.25 & 0.16667\\
$\overline{\Pi}_{2,1}$[6] & 0 &  0 & 0 & 0.25 & 0.25 & 0.33333\\
$\overline{\Pi}_{2,2}$[6] & 0 &  0 & 0 & 0.25 & 0.25 & 0.33333\\
$\overline{\Pi}_{2,3}$[2] & 0 &  0 & 0 & 0.25 & 0.25 & 0\\
$\overline{\Pi}_{2,4}$[12] & 0 &  0 & 0 & 0.25 & 0.25 & 0.16667\\
$\overline{\Pi}_{2,4}$[6] & 0 &  0 & -0.33333 & 0.25 & 0.45833 & 0\\
$\overline{\Pi}_{2,5}$[12] & 0 &  0 & -0.33333 & 0.25 & 0.33333 & 0.33333\\
$\overline{\Pi}_{2,6}$[6] & 0 & -0.33333 &  0.33333 & 0.25 & 0.16667 & 0.33333\\
$\overline{\Pi}_{2,7}$[12] & 0 &  0 & 0 & 0.25 & 0.25000 & 0.5\\
$\overline{\Pi}_{2,8}$[12] & 0 &  0 & 0 & 0.25 & 0.1667 & 0.33333\\
$\overline{\Pi}_{3,1}$[12] & 0 & 0 & 0.33333 & 0.125 & -0.08333 & 0.16667\\
$\overline{\Pi}_{3,1}$[2]  & 0 & 0 & 0 & 0.125 & 0.25 & 0.5\\
$\overline{\Pi}_{3,1}$[2]  & 0 & 0 & 0 & 0.125 & 0.25 & 0.5\\
$\overline{\Pi}_{3,2}$[24] & 0 &  0 & 0 & 0.125 & -0.04167 & 0\\
$\overline{\Pi}_{3,3}$[6] & 0 &  0 & 0 & 0.125 & -0.08333 & 0.16667 \\
$\overline{\Pi}_{3,3}$[6] & 0 &  0 & 0 & 0.125 & -0.08333 & -0.16667\\
$\overline{\Pi}_{4,1}$[4] & 0 &  0 & 0 & 0.0625 &  0 & 0.5\\
$\overline{\Pi}_{4,2}$[12] & 0 & 0 & -0.33333 & 0.0625 & -0.16667 & -0.16667\\
$\overline{\Pi}_{4,2}$[12] & 0 & 0 & 0 & 0.0625 & 0 & 0\\
$\overline{\Pi}_{4,3}$[6] & 0 &  0.33333 &  0.33333 & 0.0625 & -0.16667 & 0\\
\end{tabular}
\end{ruledtabular}
\end{table*}

The maximum range over which the correlation function of an SQS mimics that of
a random alloy can be increased by increasing the supercell size. As the size
of the SQS increases, the probability of finding configurations that mimic
random alloys over a wider coordination range increases accordingly. The search
algorithm used in this work consists of enumerating every possible supercell of
a given volume and for each supercell, enumerating every possible atomic
configuration. For each configuration, the correlation functions of different
figures, i.e. points, pairs, and triplets, are calculated. To save time, the
calculation of the correlations is stopped as soon as one of them does not
match the random state value. This algorithm becomes prohibitively expensive
very rapidly. The generation of a larger SQS could be accomplished by using a
Monte-Carlo-like scheme (e.g., \citet{1997Abr}), but this is beyond the scope
of present work. In fact, the authors could generate a 32-atom SQS's, and the
average total energy difference between 16-atom SQS's and 32-atom SQS's in the
Cd-Mg system was around 2 meV per atom. The authors maintain a focus on 16-atom
SQS, because this size represents a good compromise between accuracy and the
computational requirements associated with the necessary first-principles
calculations.

It is also important to note that finding a good hcp SQS is more difficult than
finding an SQS of cubic structures with the same range of matching correlations
due to the fact that, for a given range of correlations, there are more
symmetrically distinct correlations to match. Additionally, the lower symmetry
of the hcp structure implies that there are also many more candidate
configurations to search through in order to find a satisfactory SQS. Thus, the
number of distinct supercells is larger and the number of symmetrically
distinct atomic configurations is larger, in comparison to fcc or bcc lattices.

In order to verify the proposed 16-atom SQS's are adequate for the simulation
of hcp random solutions, the authors calculated other SQS's at 75 at.\% which
have random-like pair correlations up to the third nearest-neighbor but that
have slightly different correlations for the fourth nearest-neighbor. The pair
correlation function at 75 at.\% of a truly random solution would be
$(2\times0.75-1)^2=0.25$ and therefore the four SQS's in \tab{tbl:Correlation2}
are worse than the one used in the present work. These structures were applied
to the  Cd 25 at.\%-Mg 75 at.\% system and, as can be seen in
\tab{tbl:Correlation2}, the associated energy differences are negligible. This
is due to the fact that the energetics of this system are dominated by
short-range interactions. Thus, as long as the most important pair correlations
(up to the third nearest-neighbors in hcp structure with ideal c/a ratio) are
satisfied, the SQS's can successfully be applied to acquire properties of
random solutions in which short-range interactions dominate.

\begin{table}
\fontsize{8}{8pt}\selectfont \caption{\label{tbl:Correlation2}Pair correlation
functions up to the fifth and the calculated total energies of other 16 atoms
sqs's for Cd$_{0.25}$Mg$_{0.75}$ are enumerated to be compared with the one
used in this work (SQS-16). The total energies are given in the unit,
$eV/atom$.}
\begin{ruledtabular}
\begin{tabular}{cccccc}
                        &    a    &    b    &    c    &    d    & SQS-16 \\
\hline
 $\overline{\Pi}_{2,1}$[6]  &  0.25   &  0.25   &  0.25   &  0.25   &  0.25   \\
 $\overline{\Pi}_{2,1}$[6]  &  0.25   &  0.25   &  0.25   &  0.25   &  0.25   \\
 $\overline{\Pi}_{2,2}$[6]  &  0.25   &  0.25   &  0.25   &  0.25   &  0.25   \\
 $\overline{\Pi}_{2,3}$[2]  &  0.25   &  0.25   &  0.25   &  0.25   &  0.25   \\
 $\overline{\Pi}_{2,4}$[12] & 0.20833 & 0.16667 & 0.16667 & 0.08333 &  0.25   \\
 $\overline{\Pi}_{2,4}$[6]  &   0.5   &   0.5   &   0.5   & 0.16667 & 0.45833 \\
 $\overline{\Pi}_{2,5}$[12] &   0.5   & 0.16667 & 0.33333 & 0.33333 & 0.33333 \\
 \hline
   Symmetry    & \multirow{2}{*}{-1.3864} & \multirow{2}{*}{-1.3882} %
               & \multirow{2}{*}{-1.3886} & \multirow{2}{*}{-1.3886} %
               & \multirow{2}{*}{-1.3869}  \\
   Preserved   &                  &         &         &         &         \\
   Fully       & \multirow{2}{*}{-1.3874}   & \multirow{2}{*}{-1.3887} %
               & \multirow{2}{*}{-1.3889} & \multirow{2}{*}{-1.3893} %
               & \multirow{2}{*}{-1.3883} \\
   Relaxed     &                  &         &         &         &         \\
\end{tabular}
\end{ruledtabular}
\end{table}

\section{First-Principles Methodology}
\label{sec:First-P-Method}

The selected hcp SQS-16 structures were used as geometrical input for the
first-principles calculations. The Vienna \emph{Ab initio} Simulation Package
(VASP)\cite{1996Kre} was used to perform the Density Functional Theory (DFT)
electronic structure calculations. The projector augmented wave (PAW) method
\cite{1999Kre} was chosen and the general gradient approximation (GGA)
\cite{1992Per} was used to take into account exchange and correlation
contributions to the hamiltonian of the ion-electron system. A constant energy
cutoff of 350 eV was used for all the structures, with 5,000 {\em k}-points per
reciprocal atom based on the Monkhorst-Pack scheme for the Brillouin-zone
integrations. The k-point meshes were centered at the $\Gamma$ point. The
convergence criterion for the calculations was 10 meV with respect to the 16
atoms. Spin-polarization was not taken into account. The generated SQS's were
either fully relaxed, or relaxed without allowing local ion relaxations, i.e.,
only volume and c/a ratio were optimized. As will be seen below, the full
relaxation caused some of the SQS's to lose the original hcp symmetry.

\section{Results and discussions}
\label{sec:Results-discuss}

\subsection{Analysis of Relaxed Structures}
\label{sub:Analysi-Relaxed-Structu}%%

The symmetry of the resulting SQS was checked using the PLATON\cite{2003Spe}
code before and after the relaxations. Both SQS's have the lowest symmetry of
P1, although all the atoms are sitting on the lattice sites of hcp. The
procedure was verified by checking the symmetries of the generated
\emph{unrelaxed} SQS. Once all the sites in the SQS were substituted with one
single atomic species, PLATON identified SQS's as perfect hcp structures. All
the atoms of the initial structures are on their \emph{exact} hcp lattice
sites. However, upon relaxation the atoms may be displaced from these ideal
positions. According to the definition of an hcp random solution, all the
atoms, in this case two different type of atoms, should be at the hcp lattice
points ---within a certain tolerance--- even after the structure has been fully
relaxed. The default tolerance of detecting the symmetry of the relaxed
structures allowed the atoms to deviate from their original lattice sites by up
to 20\%.

In principle, relaxations should be performed with respect to the degrees of
freedom consistent with the initial symmetry of any given configuration. In the
particular case of the hcp SQS's, local relaxations may in some cases be so
large that the character of the underlying parent lattice is lost. However,
within the CALPHAD methodology, one has to define the Gibbs energy of a phase
throughout the entire composition range, regardless of whether the structure is
stable or not. In these cases, it is necessary to constrain the relaxations so
that they are consistent with the lattice vectors and atom positions of an hcp
lattice. Obviously, the energetic contributions due to local relaxations are
not considered in this case. The results of these constrained relaxations can
therefore be directly compared to those calculations using the CPA. In most
cases, local relaxations were not significant. However, in a few instances, it
was found that the structure was too distorted to be considered as hcp after
the full relaxation. However, this symmetry check was not sufficient to
characterize the relaxation behavior of the relaxed SQS. Furthermore, in some
of the cases it may be possible for the structure to fail the symmetry test and
still retain an hcp-like environment within the first couple of coordination
shells, implying that the energetics and other properties calculated from these
structures could be characterized as reasonable, although not optimal,
approximations of random configurations.

\subsubsection{Radial Distribution Analysis}
\label{sub:RD}%%

In order to investigate the local relaxation of the fully relaxed SQS, their
radial distribution (RD) was analyzed. Through this analysis, the bond
distribution and coordination shells were studied to determine whether the
relaxed structures maintained the local hcp-like environment they were supposed
to mimic in the first place. Additionally, this analysis permitted us to
quantify the degree of local relaxations up to the fifth coordination shells.

The RD of each of the fully relaxed structures was obtained by counting the
number of atoms within bins of $10^{-3}$\AA, up to the fifth coordination
shell. In order to eliminate high frequency noise, the raw data was scaled and
smoothed through Gaussian smearing with a characteristic distance of 0.01 \AA.
Pseudo-Voigt functions were then used to fit each of the smoothed peaks and the
goodness of fit was in part determined through the summation of the total areas
of the peaks and comparing them to the total number of atoms that were expected
within the analyzed coordination shells. The relaxation of the atoms at each
coordination shell is quantified by the width of the corresponding peak in the
fitted RD.

\begin{figure*}[htb]
\centering%
    \subfigure[~RD of $\rm Hf_{50}Zr_{50}$ ($\Delta H_{mix} \sim 0$)]{
        \includegraphics[width=3.0in]{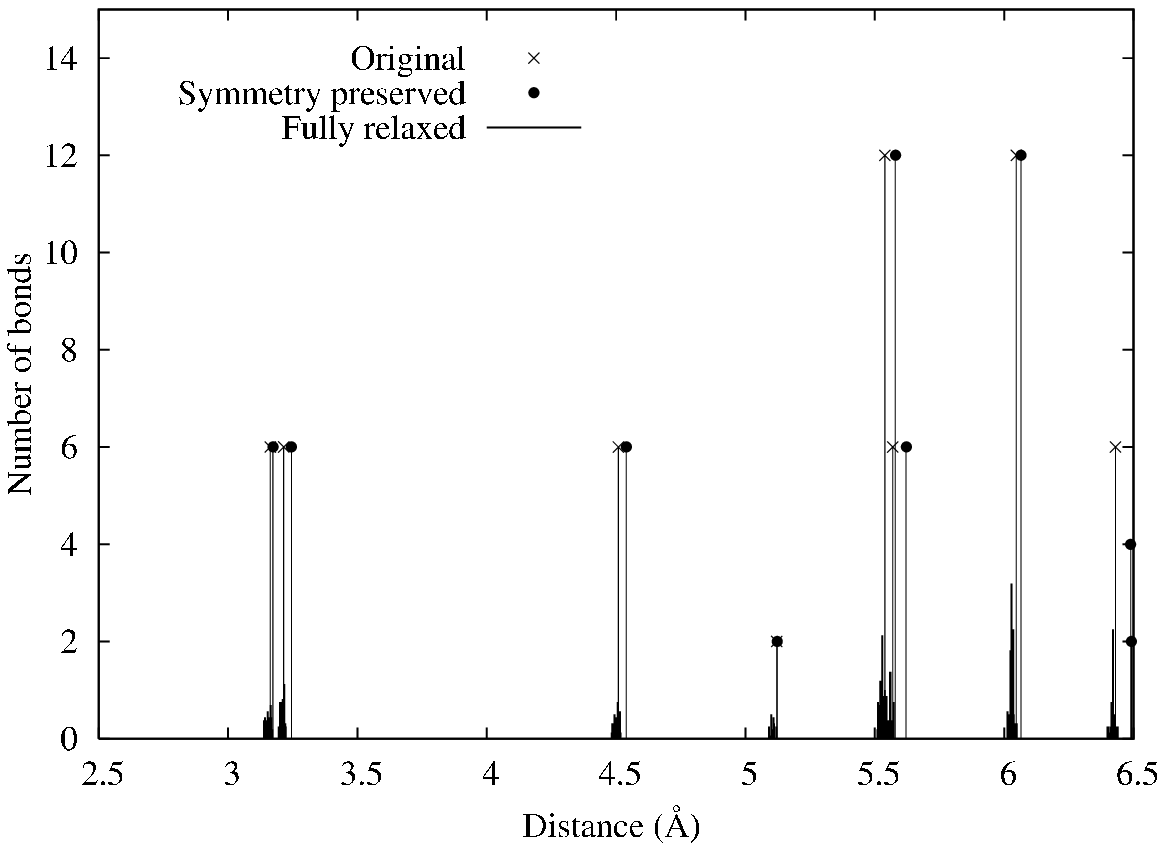}%
        \label{fig:rdfhfzr50}
    }%
    \subfigure[~Smoothed and fitted RD's of fully relaxed $\rm Hf_{50}Zr_{50}$]{
        \includegraphics[width=3.0in]{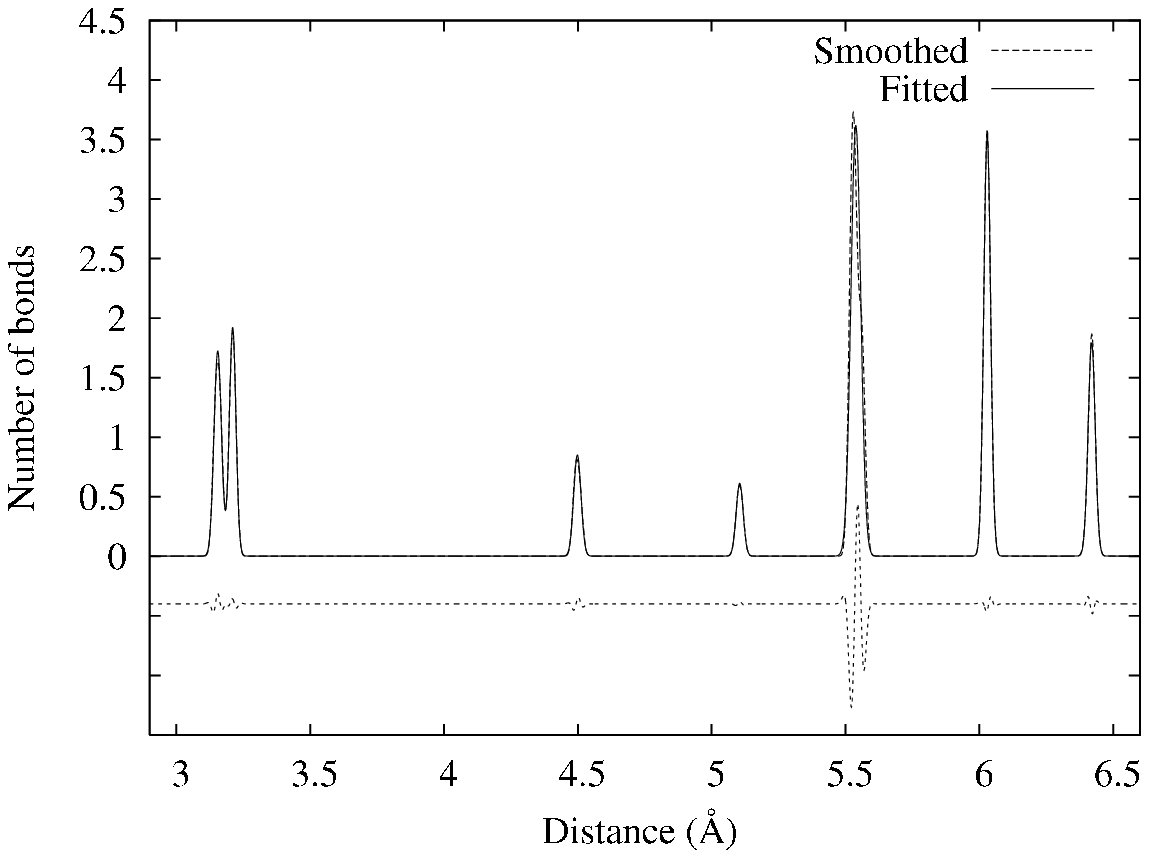}%
        \label{fig:rdhfzrfit50}
    }\\%
    \subfigure[~RD of $\rm Cd_{50}Mg_{50}$ ($\Delta H_{mix} < 0$)]{
        \includegraphics[width=3.0in]{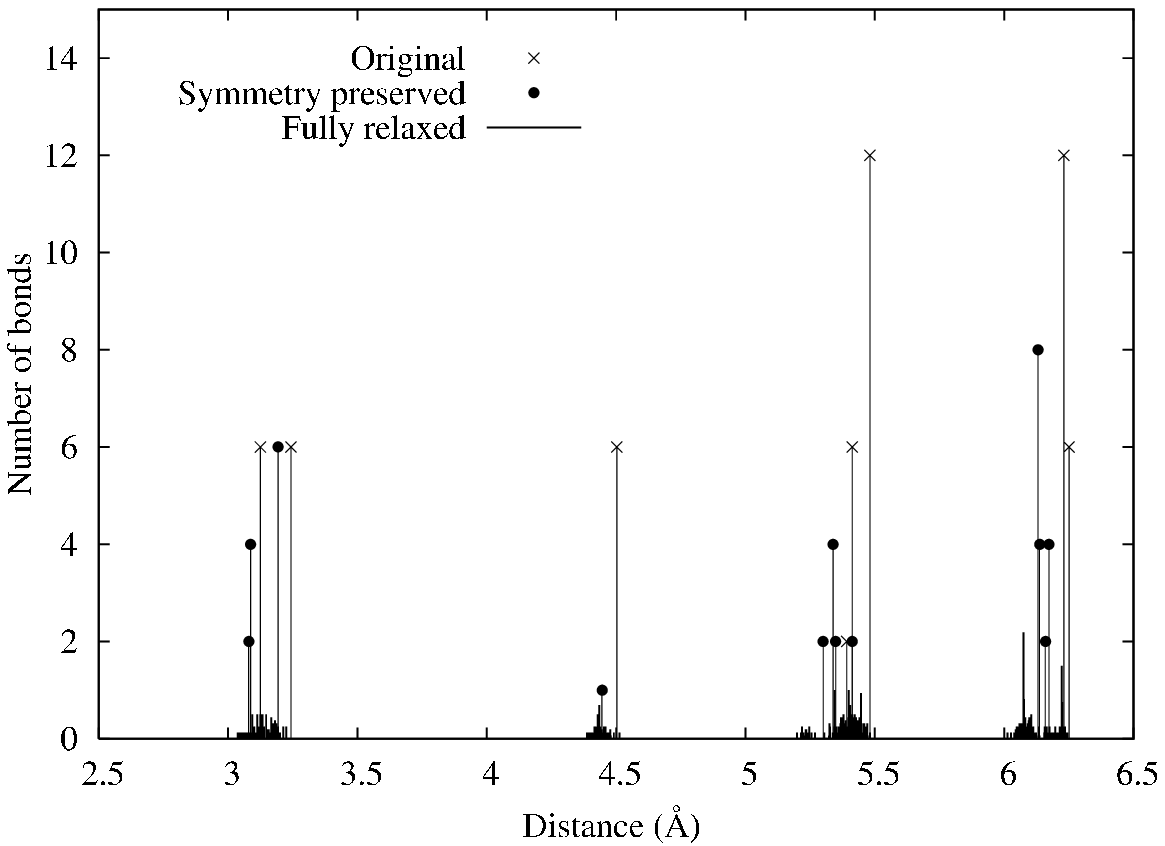}%
        \label{fig:rdfcdmg50}
    }%
    \subfigure[~Smoothed and fitted RD's of fully relaxed $\rm Cd_{50}Mg_{50}$]{
        \includegraphics[width=3.0in]{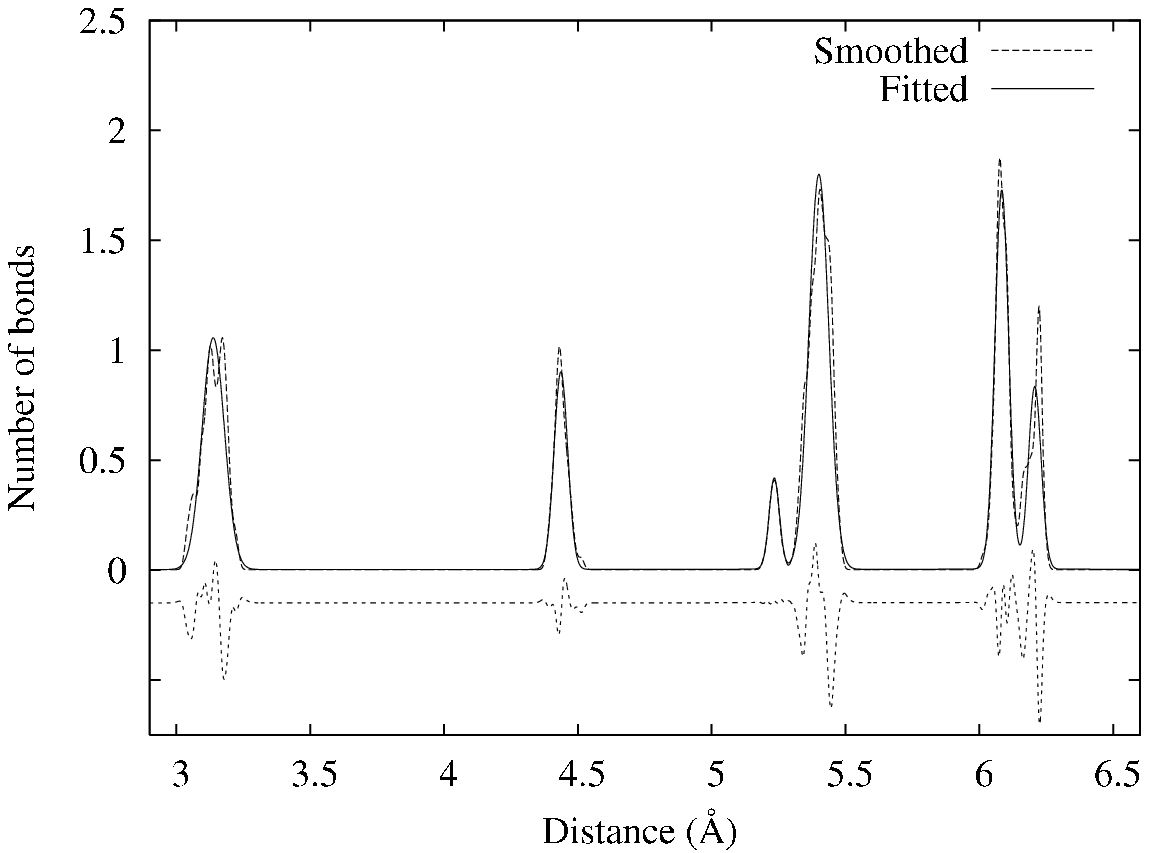}%
        \label{fig:rdfcdmgfit50}
    }\\%
    \subfigure[~RD of $\rm Mg_{50}Zr_{50}$ ($\Delta H_{mix} > 0$)]{
        \includegraphics[width=3.0in]{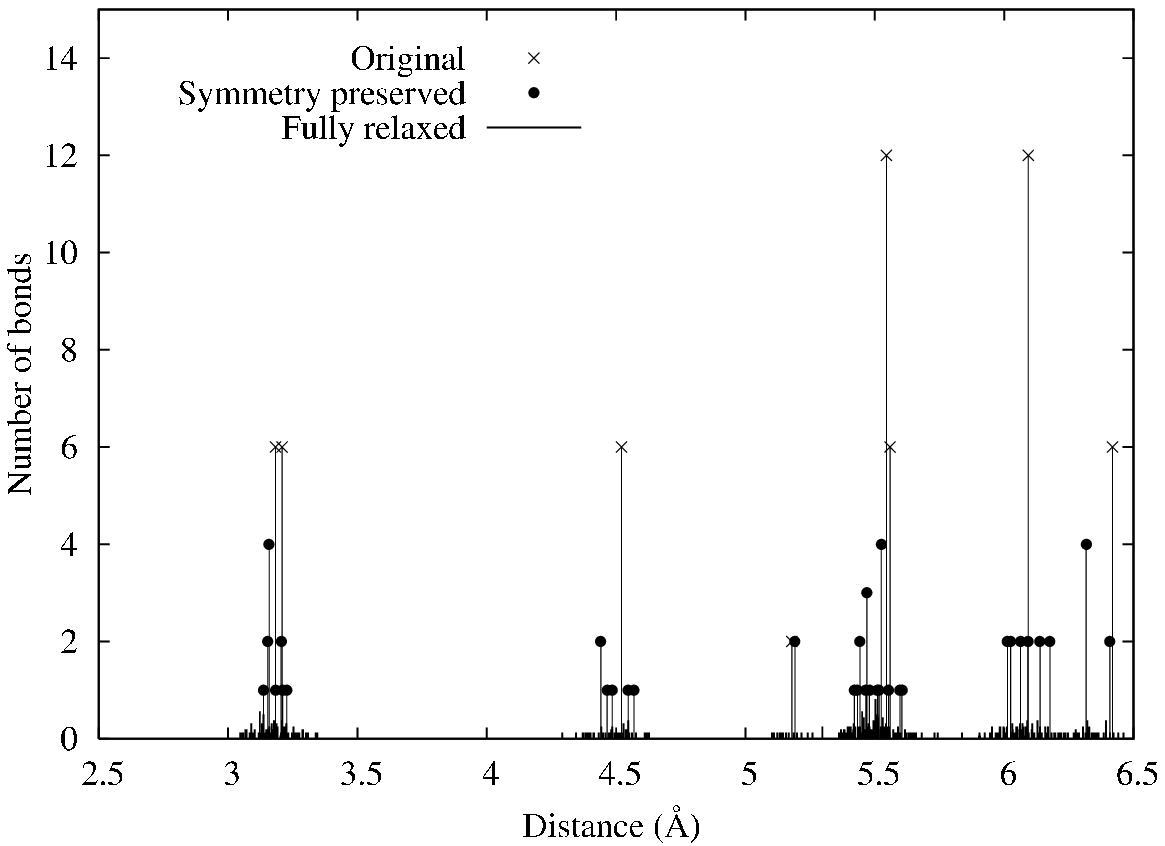}%
        \label{fig:rdfmgzr50}
    }%
    \subfigure[~Smoothed and fitted RD's of fully relaxed $\rm Mg_{50}Zr_{50}$]{
        \includegraphics[width=3.0in]{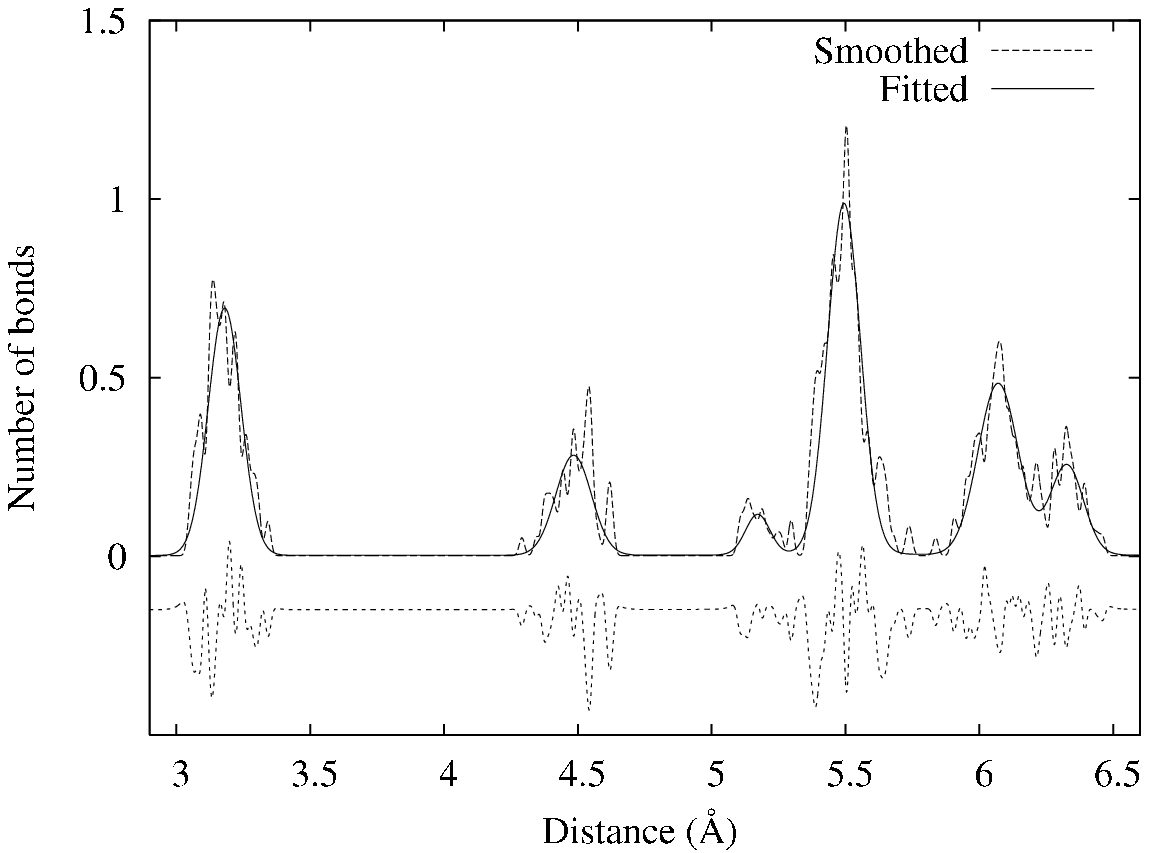}%
        \label{fig:rdfmgzrfit50}
    }%
\caption{Radial distribution analysis of selected SQS's. The dotted lines under
the smoothed and fitted curves are the error between the two curves.}
\label{fig:rdfsqs}%
\end{figure*}

The RD results of selected SQS's are given in \fig{fig:rdfsqs}. The unrelaxed,
fully relaxed, and non-locally relaxed structures are compared in each case as
well as the smoothed bond distributions and their fitted curves. These results
are representative of the RD's obtained for the seven binary systems at the
three compositions studied.

\fig{fig:rdfhfzr50} shows the RDs for the Hf-Zr SQS at the 50 at.\%
composition. As can be seen in the figure, the RDs for the unrelaxed and
non-locally relaxed SQS are almost identical, implying that in this system
Vegard's Law is closely followed. Furthermore, the RD for the fully relaxed SQS
in \fig{fig:rdhfzrfit50} shows a rather narrow distribution around each of the
the bond-lengths corresponding to the ideal or unrelaxed structure. The system
therefore needs to undergo very negligible local relaxations in order to
minimize its energy.

In the case of the Cd-Mg solution at 50 at.\% (\fig{fig:rdfcdmg50}), the RDs of
the unrelaxed and non-locally relaxed SQS are more dissimilar. Even in the
non-locally relaxed calculation, the original first coordination shell
(corresponding to the six first-nearest neighbors) has split into two different
shells (of 4 and 2 atoms) and the position of the peak is noticeably shifted.
The first two well defined coordination shells of the unrelaxed structure have
merged into a single, broad peak at 3.14\AA $~$upon full relaxation, as shown
in \fig{fig:rdfcdmgfit50}. This peak now encloses 12 first nearest neighbors.
As shown in \tab{tbl:Correlation}, $\overline{\Pi}_{2,1}$ and
$\overline{\Pi}_{2,4}$ have two different types of pairs. However, since they
have the same correlation functions, they cannot be distinguished. In
\fig{fig:rdfcdmgfit50} it is also shown how the fourth and fifth coordination
shells merge at 5.40\AA,
 enclosing 18 atoms. It can be expected that if the c/a ratio of a
relaxed structure is close to ideal and the broadening of nearby shells are
wide enough that they merge, then the structure has almost the same radial
distribution of an ideal hcp structure, albeit with a larger peak width.

\fig{fig:rdfmgzr50} shows the RD for the $\rm Mg_{50}Zr_{50}$ composition.
Among the three RD's presented in \fig{fig:rdfsqs}, this one is clearly the one
that undergoes the greatest distortion upon full relaxation. Even in the
non-locally relaxed structures there is a broad bond-length distribution around
the peaks of the unrelaxed SQS. With respect to the fully relaxed SQS, it can
be seen how the peaks for the fifth and sixth coordination shells have
practically merged. In this case, the local environment of each atom within the
SQS stops being hcp-like within the first couple of coordination shells.
Although the two end members of this binary alloy have an hcp as the stable
structure, it is evident from this figure that the SQS arrangement is unstable
and there is a tendency for the structure to distort. In this system, there is
a miscibility gap in the hcp phase up to $\sim$ 900K and the RD reflects the
tendency for the system to phase-separate.

The results from the peak fitting for all the fully relaxed SQS's are
summarized in \tab{tbl:rdf}. It should be noted that regardless of the system
and compositions, the sum of the areas under each peak should converge to a
single value, proportional to 50 atoms. For each peak, the error was quantified
as the absolute and normalized difference between the expected and actual
areas. The error reported in the table is the averaged value for all the peaks
in the RD. The broadness of the peaks in the RD is quantified through the full
width at half maximum, FWHM. In the table, the reported FWHM corresponds to the
\emph{average} FWHM observed for the coordination shells enclosing a total of
50 atoms. Note that the alloys with the smallest FWHM are Hf-Zr and Cd-Mg. As
will be seen later, Hf-Zr behaves almost ideally and Cd-Mg is a system with
rather strong attractive interactions between unlike atoms that forms ordered
hexagonal structures at the 25 and 75 at.\% compositions.

\setlength{\extrarowheight}{4pt}
\begin{table*}[htb]
\fontsize{8}{8pt}\selectfont
\caption{\label{tbl:rdf}Results of radial distribution analysis for the seven
binaries studied in this work. FWHM shows the averaged full width at half
maximum and is given in \AA. Errors indicate the difference in the number of
atoms calculated through the sum of peak areas and those expected in each
coordination shell.}
\begin{ruledtabular}
\begin{tabular}{ccccccccc}
Compositions &  & Cd-Mg & Mg-Zr & Al-Mg & Mo-Ru & Hf-Ti & Hf-Zr & Ti-Zr \\
 \hline
\multirow{3}{*}{$\rm A_{75}B_{25}$}
%          Cd-Mg               Mg-Zr              Al-Mg              Mo-Ru
& FWHM   & $0.06\pm0.01$   & $0.09\pm0.03$  & $0.08\pm0.02$  &
N/A\footnote{The radial distribution analysis of Mo 75 at.\%-Ru 25 at.\% was
not possible since it completely lost its symmetry as hcp.}
%          Hf-Ti               Hf-Zr              Ti-Zr
         & $0.11\pm0.03$   & $0.02\pm0.00$  & $0.16\pm0.05$\\
%              Cd-Mg     Mg-Zr     Al-Mg     Mo-Ru
& Error, \%  & 0.72  & 0.39  & 0.47  & N/A
%              Hf-Ti     Hf-Zr     Ti-Zr
             & 1.07  & 1.84  & 1.27 \\
%            Cd-Mg     Mg-Zr     Al-Mg     Mo-Ru
& Symmetry & PASS    & PASS    & PASS    & FAIL
%            Hf-Ti     Hf-Zr     Ti-Zr
           & PASS    & PASS    & FAIL \\

\hline \multirow{3}{*}{$\rm A_{50}B_{50}$}
%          Cd-Mg             Mg-Zr             Al-Mg             Mo-Ru
& FWHM   & $0.07\pm0.02$ & $0.15\pm0.02$ & $0.15\pm0.07$ &$0.13\pm0.01$
%          Hf-Ti               Hf-Zr              Ti-Zr
         & $0.16\pm0.02$   & $0.03\pm0.01$  & $0.19\pm0.06$ \\
%              Cd-Mg     Mg-Zr     Al-Mg    Mo-Ru
& Error, \%  & 0.30  & 1.42  & 1.28 & 1.90
%              Hf-Ti     Hf-Zr     Ti-Zr
             & 0.35  & 1.84  & 2.39 \\
%            Cd-Mg     Mg-Zr     Al-Mg     Mo-Ru
& Symmetry & PASS    & FAIL    & FAIL    & PASS
%            Hf-Ti     Hf-Zr     Ti-Zr
           & PASS    & PASS    & PASS \\

\hline \multirow{3}{*}{$\rm A_{25}B_{75}$}
%          Cd-Mg             Mg-Zr             Al-Mg             Mo-Ru
& FWHM   & $0.04\pm0.01$ & $0.09\pm0.03$ & $0.10\pm0.02$ &$0.07\pm0.02$
%          Hf-Ti               Hf-Zr              Ti-Zr
         & $0.11\pm0.06$   & $0.03\pm0.00$  & $0.13\pm0.07$ \\
%              Cd-Mg     Mg-Zr     Al-Mg     Mo-Ru
& Error, \%  & 2.05  & 1.22  & 0.26  & 1.93
%              Hf-Ti     Hf-Zr     Ti-Zr
             & 0.26  & 1.01   & 0.96 \\
%            Cd-Mg     Mg-Zr     Al-Mg     Mo-Ru
& Symmetry & PASS    & PASS    & PASS    & PASS
%            Hf-Ti     Hf-Zr     Ti-Zr
           & PASS    & PASS    & PASS \\
\end{tabular}
\end{ruledtabular}
\end{table*}

\subsubsection{Bond Length Analysis}
\label{sub:Bond-Length}%%

\setlength{\extrarowheight}{4pt}
\begin{table*}[htb]
\fontsize{8}{8pt}\selectfont
\caption{\label{tbl:bond}First nearest-neighbors average bond lengths for the
fully relaxed hcp SQS of the seven binaries studied in this work. Uncertainty
corresponds to the standard deviation of the bond length distributions.}
\begin{ruledtabular}
\begin{tabular}{ccccccccc}
Compositions & Bonds & Cd-Mg & Mg-Zr & Al-Mg & Mo-Ru & Hf-Ti & Hf-Zr & Ti-Zr \\
 \hline
%                                  Cd            Mg           Al
$\rm A_{100}B_{0}$ & A-A &  $3.07$  & $3.18$ &  $2.87$ &
%
%    Mo             Hf        Hf           Ti %%
$2.75$ & $3.13$ & $3.13$ & $2.87$ \\\hline
\multirow{3}{*}{$\rm A_{75}B_{25}$}
%           Cd-Mg            Mg-Zr            Al-Mg           Mo-Ru
& A-A     & $3.17\pm0.10$  & $3.18\pm0.03$ & $2.92\pm0.03$ &
%           Hf-Ti             Hf-Zr           Ti-Zr
          & $3.14\pm0.05$  & $3.18\pm0.03$ & $2.96\pm0.07$ \\
%            Cd-Mg           Mg-Zr           Al-Mg            Mo-Ru
& A-B     & $3.16\pm0.11$ & $3.18\pm0.05$  & $2.95\pm0.03$ & N/A
%           Hf-Ti             Hf-Zr           Ti-Zr
          & $3.10\pm0.05$  & $3.18\pm0.03$ & $3.02\pm0.07$ \\
%            Cd-Mg           Mg-Zr           Al-Mg            Mo-Ru
& B-B     & $3.18\pm0.10$ & $3.12\pm0.10$  & $2.96\pm0.03$ &
%           Hf-Ti             Hf-Zr           Ti-Zr
          & $3.09\pm0.06$  & $3.18\pm0.04$ &  $3.04\pm0.06$\\

\hline \multirow{3}{*}{$\rm A_{50}B_{50}$}
%            Cd-Mg           Mg-Zr           Al-Mg           Mo-Ru
& A-A     & $3.16\pm0.04$  & $3.16\pm0.04$ & $2.98\pm0.06$  & $2.81\pm0.08$
%           Hf-Ti             Hf-Zr           Ti-Zr
          & $3.09\pm0.06$  & $3.18\pm0.03$ & $3.00\pm0.09$ \\
%            Cd-Mg           Mg-Zr           Al-Mg            Mo-Ru
& A-B     & $3.12\pm0.04$  & $3.20\pm0.06$ & $3.02\pm0.06$  & $2.75\pm0.04$
%           Hf-Ti             Hf-Zr           Ti-Zr
          & $3.05\pm0.07$  & $3.19\pm0.03$ & $3.06\pm0.08$ \\
%            Cd-Mg           Mg-Zr           Al-Mg            Mo-Ru
& B-B     & $3.15\pm0.03$  & $3.14\pm0.08$ & $3.07\pm0.08$  & $2.75\pm0.04$
%           Hf-Ti             Hf-Zr           Ti-Zr
          & $3.00\pm0.06$  & $3.20\pm0.03$ &  $3.12\pm0.08$\\

\hline \multirow{3}{*}{$\rm A_{25}B_{75}$}
%            Cd-Mg           Mg-Zr           Al-Mg            Mo-Ru
& A-A     & $3.16\pm0.01$  & $3.15\pm0.04$ & $3.06\pm0.04$  & $2.73\pm0.04$
%           Hf-Ti             Hf-Zr           Ti-Zr
          & $3.02\pm0.05$  & $3.19\pm0.03$ & $3.09\pm0.08$ \\
%            Cd-Mg           Mg-Zr           Al-Mg            Mo-Ru
& A-B     & $3.14\pm0.02$  & $3.19\pm0.04$ & $3.08\pm0.04$  & $2.73\pm0.04$
%           Hf-Ti             Hf-Zr           Ti-Zr
          & $3.00\pm0.06$  & $3.19\pm0.03$ & $3.11\pm0.06$ \\
%            Cd-Mg           Mg-Zr           Al-Mg            Mo-Ru
& B-B     & $3.15\pm0.01$ & $3.18\pm0.04$  & $3.11\pm0.03$  & $2.71\pm0.04$
%           Hf-Ti             Hf-Zr           Ti-Zr
          & $2.95\pm0.05$  & $3.20\pm0.04$ &  $3.17\pm0.06$\\\hline
%                                  Mg            Zr           Mg
$\rm A_{0}B_{100}$ & B-B &  $3.18$  & $3.19$ &  $3.18$ &
%
%%    Ru           Ti           Zr           Zr %%
$2.68$ & $2.87$ & $3.19$ & $3.19$ \\
\end{tabular}
\end{ruledtabular}
\end{table*}

In addition to the RD analysis, we performed the bond length analysis (A-A,
B-B, and A-B) for all the relaxed SQS's. In \tab{tbl:bond} the bond lengths
corresponding to the first nearest neighbors for all the 21 SQS's are
presented. As expected, in the majority of the cases the sequence
$d_{ii}<d_{ij}<d_{jj}$ is observed throughout the composition range, where
$d_{ij}$ corresponds to the bond distance between two different atom types. The
two notable exceptions to this trend correspond to the Cd-Mg and Mg-Zr alloys.
As will be mentioned below, the Cd-Mg system tends to form rather stable
intermetallic compounds at the 25, 50 and 75 at.\% compositions, including two
hexagonal intermetallic compounds. The calculated enthalpy of mixing in this
case---shown in \fig{fig:cdmgh}---is the most negative among seven binaries
studied and the fact that the Cd-Mg bonds are shorter than Cd-Cd and Mg-Mg
seems to reflect the tendency of this system to order. In the case of the Mg-Zr
alloys, the Mg-Zr bonds are longer than Mg-Mg and Zr-Zr, suggesting that this
system has a great tendency to phase separate, as indicated by the presence of
a large hcp miscibility gap in the Mg-Zr phase diagram~\cite{Nay85}.

\subsection{Enthalpy of Mixing}
\label{sub:Enthalp-Mixing}

It is obvious that if an hcp SQS alloy is not stable with respect to local
relaxations, its properties are not accessible through experimental
measurements. However, approximate \emph{effective} properties could still be
estimated through CALPHAD modeling. In order to compare the energetics and
properties of the calculated SQS's with the available experiments or previous
thermodynamic models, only the non-locally relaxed structures were considered
whenever the SQS was identified as unstable. This effectively assumes that the
structures in question are constrained to maintain their symmetry. The total
energies of the structures under symmetry-preserving relaxations are obviously
higher since the relaxation energy is not considered. However, we can consider
these calculated thermochemical properties as an upper bound which can still be
of great use when attempting to generate thermodynamically consistent models
based on the combined first-principles/CALPHAD approach.

As mentioned earlier, obtaining thermodynamic properties of random alloys using
cluster expansion or the CPA method has some drawbacks. These methods, however,
have the advantage of calculating the properties of random alloys at arbitrary
and closely spaced concentrations. SQS's in this case are at a disadvantage
since the size of the SQS itself limits the concentrations with random-like
correlations. Nevertheless, if we can acquire the properties at these three
compositions, we can sufficiently describe the tendency of the system.
Furthermore, these SQS's can be applied directly to other binary systems
without any modifications.

The enthalpies of mixing for these alloys were calculated at the 25, 50, and 75
at.\% concentrations through the expression:

\begin{equation}\label{eqn:pure}
\Delta H(A_{1-x}B_x)=E(A_{1-x}B_x)-(1-x)E(A)-xE(B)
\end{equation}

\noindent where $E(A)$ and $E(B)$ are the reference energies of the pure
components in their hcp ground state.

In the following sections, the generated SQS's are tested by calculating the
crystallographic, thermodynamic and electronic properties of hcp random
solutions in seven binary systems, Cd-Mg, Mg-Zr, Al-Mg, Mo-Ru, Hf-Ti, Hf-Zr,
and Ti-Zr. The results of the calculations are then compared with existing
experimental information as well as previous calculations.

\subsection{Cd-Mg}
\label{sub:Cd-Mg}%%

In the Cd-Mg system, both elements have the same valence and almost the same
atomic volumes. Consequently, there is a wide hcp solid solution range as well
as order/disorder transitions in the central, low temperature region of the
phase diagram. In fact, at the 25 and 75 at.\% compositions there are ordered
intermetallic phases with hexagonal symmetries.

\begin{figure*}[htb]
\centering %
    \subfigure[~Calculated enthalpy of mixing for the disordered hcp
phase in the Cd-Mg system with SQS at T=0K, Cluster Variation Method
(CVM)\cite{1993Ast} at T=900K, and experiment\cite{1963Hul} at
T=543K]{
        \label{fig:cdmgh} %
        \includegraphics[width=3.0in]{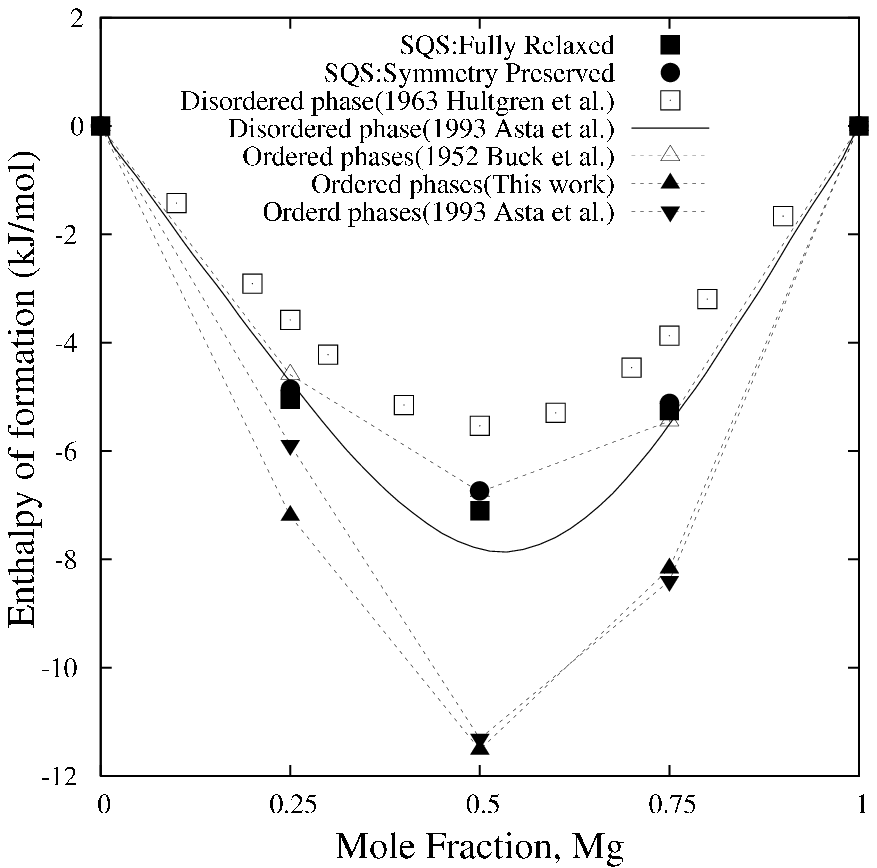}
    }
    \subfigure[~Calculated lattice parameters of the Cd-Mg system compared with
experimental data\cite{1940Hum,1957Von,1959Har}]{
        \label{fig:cdmglattice} %
        \includegraphics[width=3.0in]{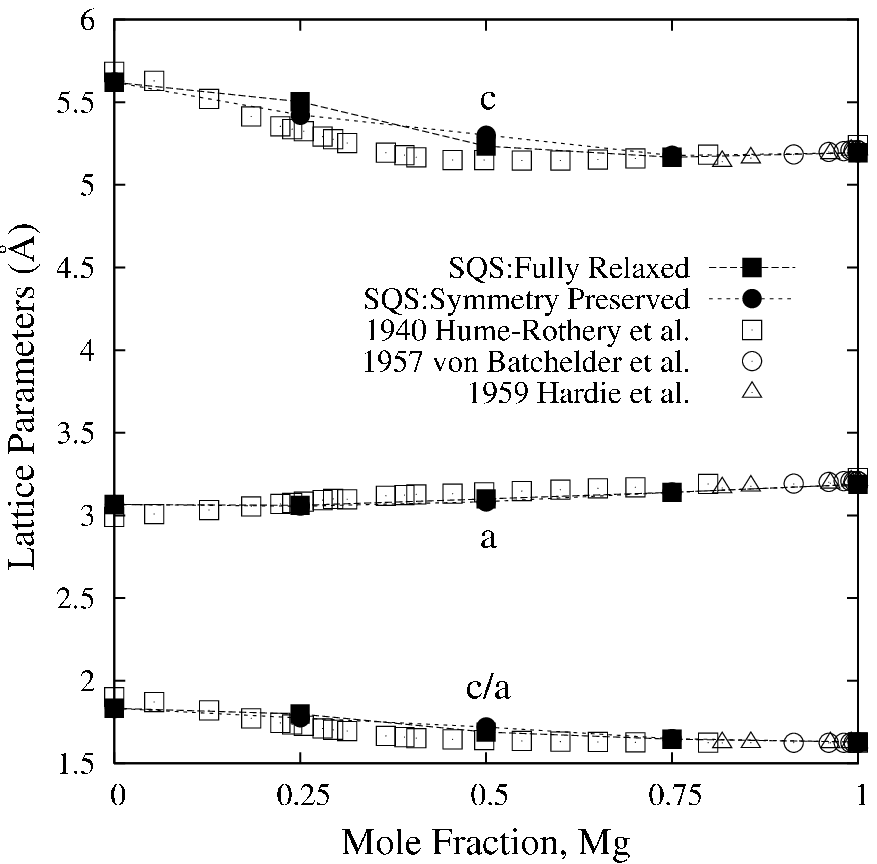}
    }
\caption{%
Calculated and experimental results of mixing enthalpy and lattice
parameters for the Cd-Mg system
}%
\label{fig:cdmg}
\end{figure*}

\fig{fig:cdmgh} compares the enthalpy of mixing calculated from the fully
relaxed and symmetry preserved SQS with the results from cluster
expansion\cite{1993Ast}. The results by \citet{1993Ast} at 900K are presented
for comparison since it is to be expected that these values would be rather
close to the calculated enthalpy of completely disordered structures. The
previous and current calculations are also compared with the experimental
measurements as reported in \citet{1963Hul} at 543K. The first thing to note
from \fig{fig:cdmgh} is that the fully relaxed and symmetry preserved
calculations are very close in energy, implying negligible local relaxation.
Additionally, the present calculations are remarkably close ($\sim$1 kJ/mol) to
the experimental measurements. By comparing the SQS enthalpy of mixing with the
results from the cluster expansion calculations~\cite{1993Ast}, it is obvious
that the former is, at least in this case, more capable of reproducing the
experimental measurements.

Formation enthalpies of the three ordered phases in the Cd-Mg system, $\rm
Cd_3Mg$, $\rm CdMg$, and $\rm CdMg_3$ are also presented. The measurements from
\citet{1963Hul} deviate from the calculated results from \citet{1993Ast} and
this work. Cd and Mg are known as very active elements and it is likely that
reaction with oxygen present during the measurements may have introduced some
systematic errors. Furthermore, the measurements were conducted at relatively
low temperatures, making it difficult for the systems to equilibrate.
Nevertheless, experiments and calculations agree that these three compounds
constitute the ground state of the Cd-Mg system.

\fig{fig:cdmglattice} also shows that the present calculations are able to
reproduce the available measurements on the variation of the lattice parameters
of hcp Cd-Mg alloys with composition, as well as the deviation of these
parameters from Vegard's Law. This deviation is mainly related to the rather
large difference in c/a ratio between Cd and Mg. The c/a ratio of Cd is one of
the largest ones of all the stable hcp structures in the periodic table.

\subsection{Mg-Zr}\label{sub:Mg-Zr}%%

\begin{figure}[hbt]%
\centerline%
{\includegraphics[width=3.0in]{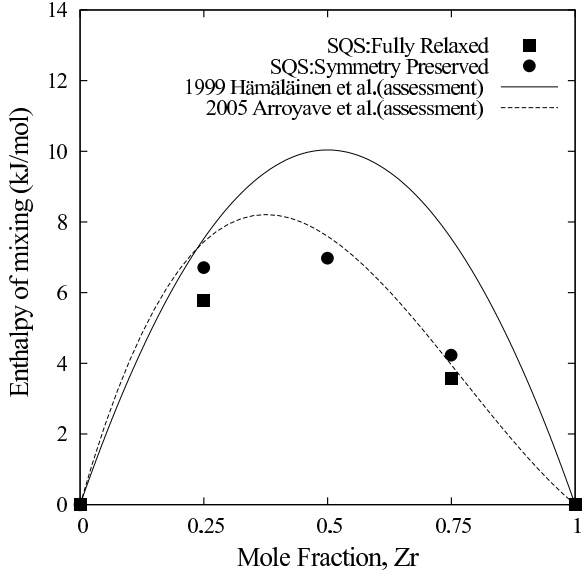}}%
\caption{%
Calculated enthalpy of mixing in the Mg-Zr system compared with a
previous thermodynamic assessment\cite{1999Hal}. Both reference states are the
hcp structure.
}%
\label{fig:mgzrh}%
\end{figure}

The Mg-Zr system is important due to the grain refining effects of Zr in
magnesium alloys. According to the assessment of the available experimental
data by \citet{Nay85}, the Mg-Zr system shows very little solubility in the
three solution phases, bcc, hcp and liquid. In fact, the low temperature hcp
phase exhibits a broad miscibility gap up to 923K, corresponding to the
peritectic reaction $hcp + liquid \to hcp$~\cite{Nay85}.

Our calculations yielded a positive enthalpy of mixing, confirming the trends
derived from the thermodynamic model developed by \citet{1999Hal}. In the case
of the full relaxation, however, it was observed that the $\rm Mg_{50}Zr_{50}$
SQS was unstable with respect to local relaxations. The instability at this
composition and the large, positive enthalpy of mixing indicate that the system
has a strong tendency to phase-separate. By comparing the fully relaxed and the
non-locally relaxed structures, we estimate that the local relaxation energy
lowers the mixing enthalpy of the random hcp SQS by about 2 kJ/mol in this
system.

\fig{fig:mgzrh} shows the calculated mixing enthalpy for the Mg-Zr hcp SQS with
no local relaxations, as well as the mixing enthalpy calculated from the
thermodynamic model by \citet{1999Hal}, which was fitted only through phase
diagram data. It is therefore remarkable that the maximum difference between
the CALPHAD model and the present hcp SQS calculations is $\sim$3 kJ/mol. The
CALPHAD model, however, does not correctly describe the asymmetry of the mixing
enthalpy indicated by the first-principles calculations. The results of the hcp
SQS calculations for the Mg-Zr system have recently been used to obtain a
better thermodynamic description of the Mg-Zr system~\cite{2005Arro} and, as
can be seen in the figure, this description is better at describing the trends
in the calculated enthalpy of mixing.

\subsection{Al-Mg}
\label{sub:Al-Mg}%%

As one of the most important industrial alloys, the Al-Mg system has been
studied extensively recently\cite{1997Lia,1998Ans,2005Zho}. This system has two
eutectic reactions and shows solubility within both the fcc and hcp phases.
However, the solubility ranges are not wide enough so there is only limited
experimental information for the properties of the hcp phase. The maximum
equilibrium solubility of Al in the Mg-rich hcp phase is around 12 at.\%.

\begin{figure*}[htb]
\centering %
    \subfigure[~Calculated enthalpy of mixing for the hcp phase in the
Al-Mg system compared with assessed data\cite{1997Lia,1998Ans,2005Zho}.
Reference states are hcp for both elements.]{
        \label{fig:almgh}%
        \includegraphics[width=3.0in]{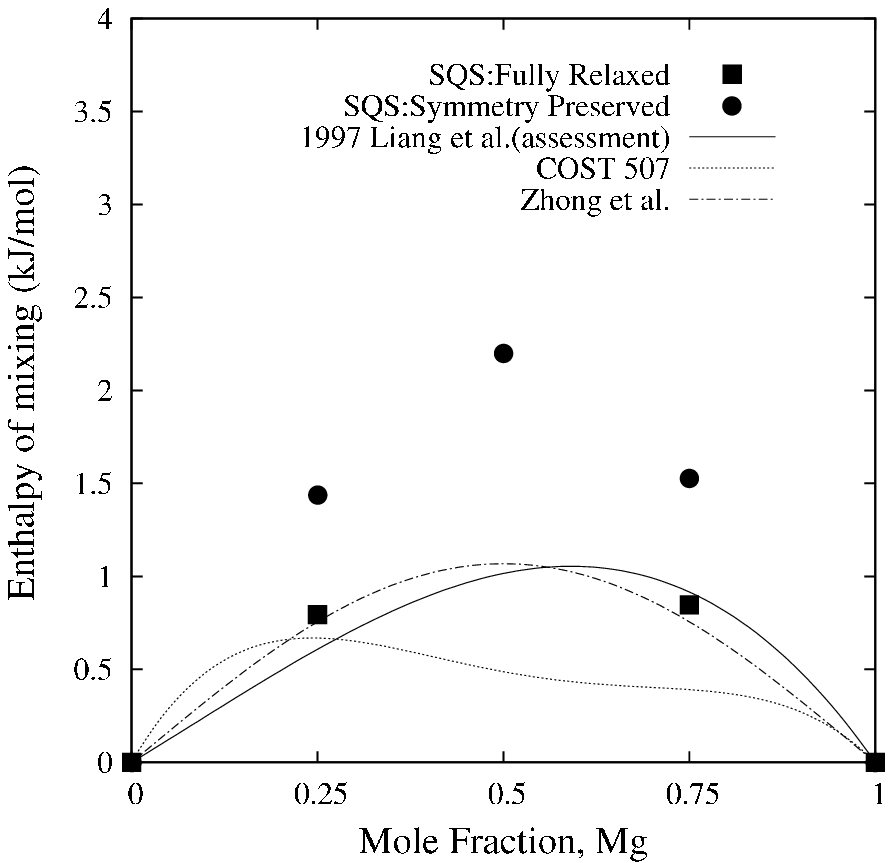}
    }
    \subfigure[~Calculated lattice parameters of the hcp phase in the Al-Mg
system compared with experimental
data\cite{1964Luo,1941Hum,1942Ray,1950Bus,1957Von,1959Har}]{
        \label{fig:almglattice}%
        \includegraphics[width=3.0in]{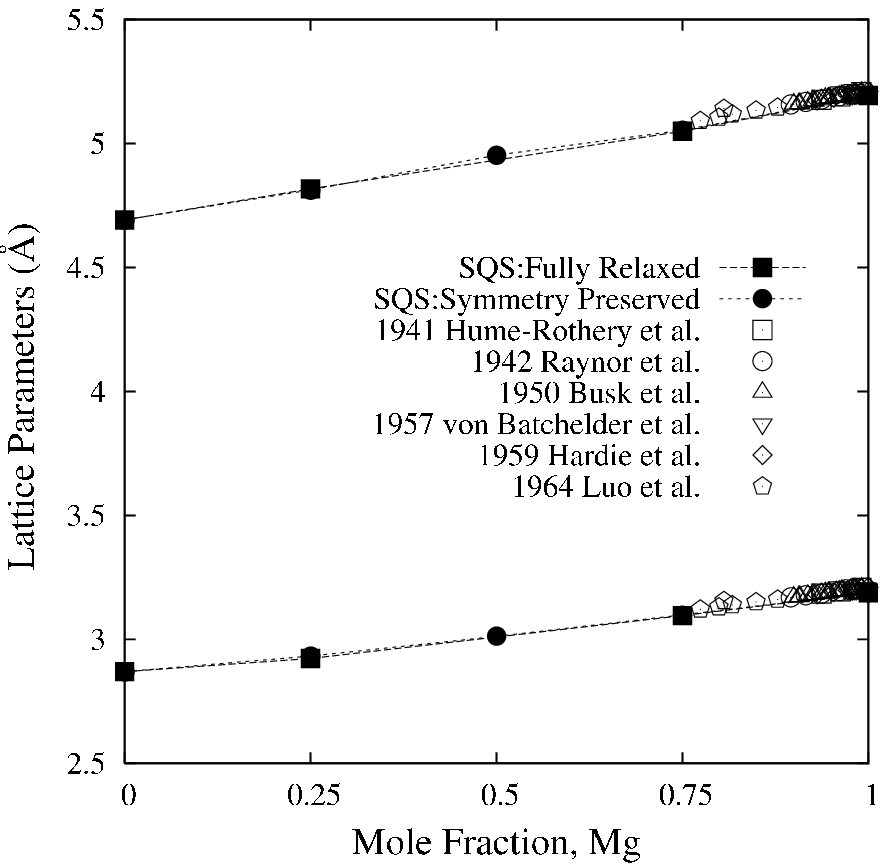}
    }
\caption{%
Calculated and experimental results of mixing enthalpy and lattice
parameters for the Al-Mg system
}%
\label{fig:almg}
\end{figure*}

In \fig{fig:almgh} the calculated enthalpy of mixing is slightly positive. The
fully relaxed calculations show that the SQS with the 50 at. \% composition was
unstable with respect to local relaxations. This can be explained by the strong
interaction between Al and Mg, as evident from the tendency of this system to
form intermetallic compounds at the middle of the phase diagram, such as
$\beta$-$\rm Al_{140}Mg_{89}$, $\gamma$-$\rm Al_{12}Mg_{17}$, and
$\varepsilon$-$\rm Al_{30}Mg_{23}$. At the 25 and 75 at.\% compositions the
SQS's were stable with respect to local relaxations because both elements have
a close-packed structure. Furthermore, at these compositions either the fcc or
hcp phase take part in equilibria with some other (intermetallic) phase.
\fig{fig:almgh} shows that the present fully relaxed calculations are in
excellent agreement with the most recent CALPHAD
assessments\cite{1997Lia,1998Ans}. Note also that in this case, and contrary to
what is observed in the Cd-Mg binary, the energy change associated with local
relaxation is not negligible, although it is still within $\sim$1 kJ/mol.

Additionally, the calculated lattice parameters agree very well with the
experimental measurements of Mg-rich hcp alloys, as can be seen in
\fig{fig:almglattice}. It is important to note that the lattice parameter
measurements of metastable hcp alloys from \citet{1964Luo}(77.4 and 87.8 Mg
at.\%) are lying on the extrapolated line between the 75 at.\% SQS and the pure
Mg calculations. This is another example of how SQS's can be successfully used
in calculating the properties of an hcp solid solution system with narrow
solubility range and mixed with non-hcp elements, even in the metastable
regions of the phase diagram.

\subsection{Mo-Ru}
\label{sub:Mo-Ru}%%

The Mo-Ru system shows a wide solubility range within both the bcc and hcp
sides of the phase  diagram. In the Ru-rich side, the maximum solubility of Mo
in the hcp-Ru matrix is up to 50 at.\%. The calculations at $\rm
Mo_{25}Ru_{75}$ and $\rm Mo_{50}Ru_{50}$ retained the original hcp symmetry but
$\rm Mo_{75}Ru_{25}$ did not. The instability of the Mo-rich SQS is not
surprising since the Mo-rich bcc region is stable over a wide region of the
phase diagram. As shown in \citet{2003Wan}, elements whose ground state is bcc
are not stable in an hcp lattice and viceversa (bcc Ti, Zr and Hf are only
stabilized at high temperature due to anharmonic effects). Thus hcp
compositions close to the bcc-side would be dynamically unstable and would have
a very large driving force to decrease their energy by transforming to bcc.

Recently, \citet{2005Kis} calculated the enthalpy of mixing for disordered hcp
Mo-Ru alloys through the CPA in which relaxation energies were estimated by
locally relaxing selected multi-site atomic arrangements. Enthalpy of formation
for hcp solutions were calculated from Eqn.\ref{eqn:moru} shown below. The
enthalpy of mixing of the disordered hcp phase can be evaluated accordingly
based on the so-called lattice stability\cite{1998Sau},
$E^{bcc}(Mo)-E^{hcp}(Mo)$.

\begin{widetext}
\begin{eqnarray}\label{eqn:moru}
\Delta H_f(Mo_{1-x}Ru_x) =&& E^{hcp}(Mo_{1-x}Ru_x) -(1-x)E^{bcc}(Mo)-xE^{hcp}(Ru) \nonumber \\
=&& E^{hcp}(Mo_{1-x}Ru_x)-(1-x)E^{hcp}(Mo) -xE^{hcp}(Ru) \nonumber\\
&&-(1-x)E^{bcc}(Mo)+(1-x)E^{hcp}(Mo)\nonumber \\
=&& H^{hcp}_{mix}(Mo_{1-x}Ru_x) -(1-x)[E^{bcc}(Mo)-E^{hcp}(Mo)]
\end{eqnarray}
\end{widetext}

Usually, structural energy differences (or lattice stability) between
first-principles calculations and CALPHAD show quite good agreement. However,
for some transition elements, the  disagreement between the two approaches is
quite significant\cite{2004Wan}. Mo is one such case, with the structural
energy difference between bcc and hcp from first-principles calculations and
the CALPHAD approach differing by over 30 kJ/mol. After a rather extensive
analysis, \citet{2005Kis} arrived at the conclusion that in order to reproduce
enthalpy values close enough to the available experimental data\cite{1988Kle}
the CALPHAD lattice stability (11.55 kJ/mol) needed to be used for the value of
the $\rm bcc \to hcp$ promotion energy.

\begin{figure*}[htb]
\centering %%
    \subfigure[~Enthalpy of formation of hcp phase in the Mo-Ru system from SQS's
(this work) and CPA\cite{2005Kis}. Total energy of hcp Mo is obtained from
first-principles calculations in both cases.]{
        \label{fig:moruhabinit}
        \includegraphics[width=3.0in]{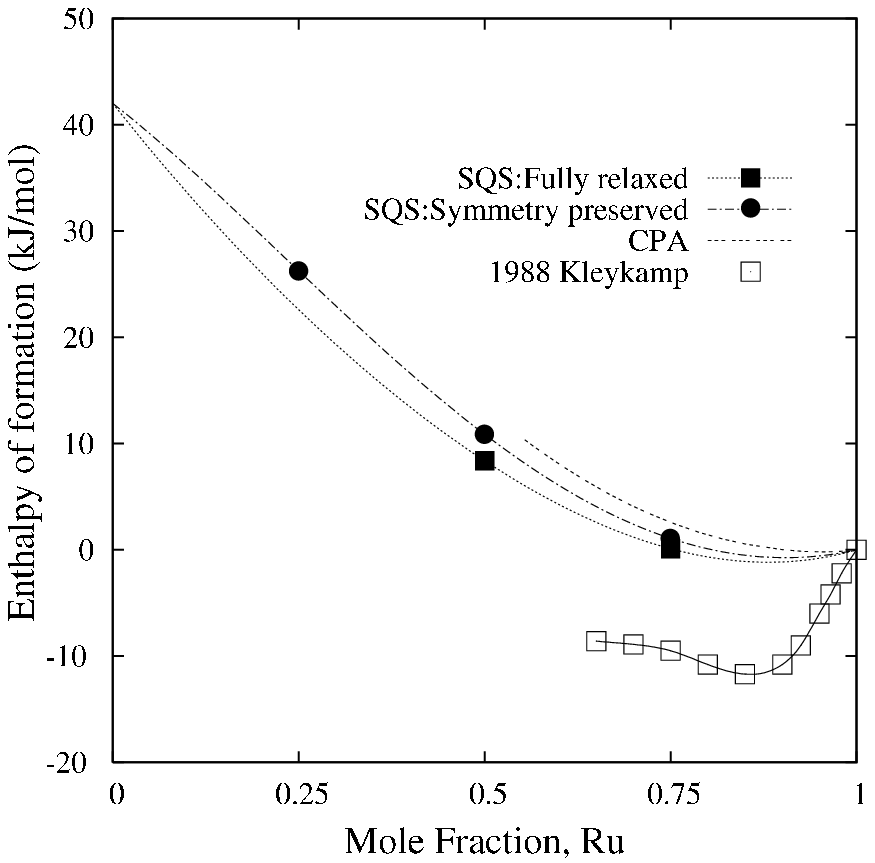}
    }
    \subfigure[~Enthalpy of formation of hcp phase in the Mo-Ru system from SQS's
and CPA. Total energy of hcp Mo is derived from the SGTE (Scientific Group
Thermodata Europe) lattice stability]{
        \label{fig:moruhsgte} %%
        \includegraphics[width=3.0in]{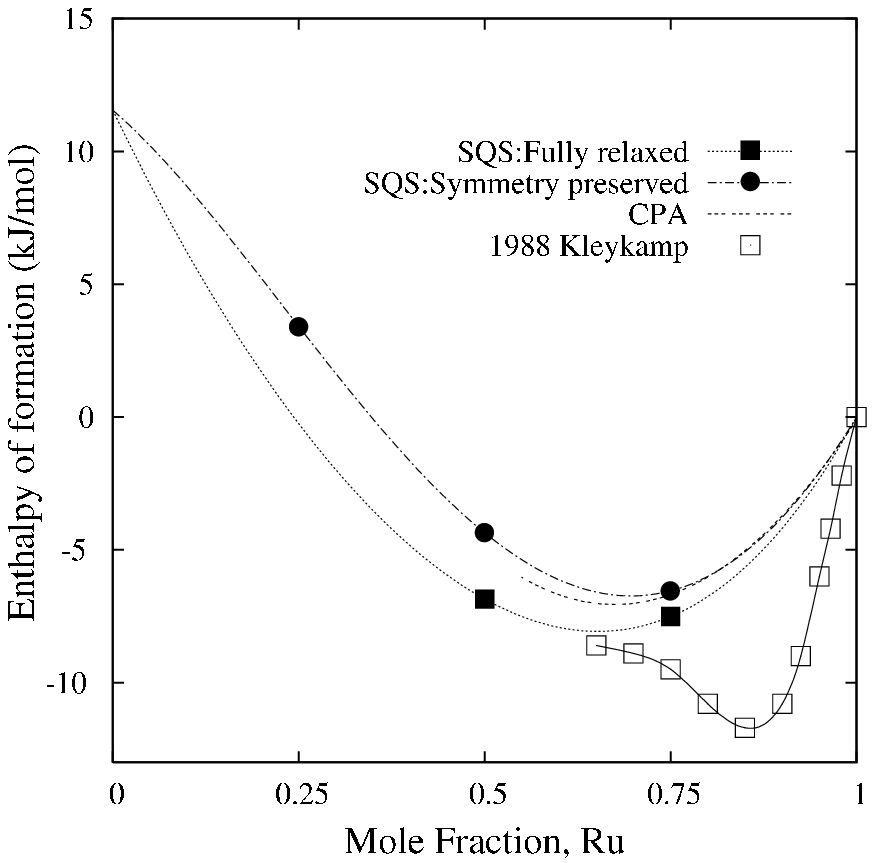}
    }
\caption{%
Enthalpy of formation of the Mo-Ru system with both first principles
and CALPHAD lattice stabilities. Reference states are bcc for Mo and hcp for
Ru.
}%
\label{fig:moru}
\end{figure*}

The SQS and CPA calculations are compared with the experimental measurements in
\fig{fig:moru}. On the assumption that the experimental measurements by
\citet{1988Kle} are correct, the derived enthalpy of formation of the hcp Mo-Ru
system from the first-principles calculated lattice stability with the SQS and
CPA approach in \fig{fig:moruhabinit} cannot reproduce the experimental
observation at all since the first-principles $\rm bcc\to hcp$ lattice
stability for Mo is 42 kJ/mol. Given this lattice stability, the only way in
which the first-principles calculations within both the SQS and CPA approaches
would match the experimental results would be for the calculated enthalpy of
mixing to be very negative, which is not the case. In fact, as can be seen in
\fig{fig:moruhabinit}, the SQS and CPA calculations are very close to each
other.

On the other hand, the enthalpy of formation derived from the CALPHAD lattice
stability in \fig{fig:moruhsgte} shows a better agreement than that from the
first-principles lattice stability. It is important to note that the CALPHAD
lattice stability was obtained through the extrapolation of phase boundaries in
phase diagrams with Mo and stable hcp elements and, therefore, are empirical.
The reason why such an empirical approach would yield a much better agreement
with experimental data is still the source of intense debate within the CALPHAD
community and has not been resolved as of now. The main conclusion of this
section, however, is that the SQS's were able to reproduce the thermodynamic
properties of hcp alloys as good as or better than the CPA method while at the
same time allowing for the ion positions to locally relax around their
equilibrium positions.

\subsection{IVA Transition Metal Alloys}\label{sub:tmetals}

\begin{figure*}[htb]
\centering%
    \subfigure[~Calculated enthalpy of mixing for the hcp phase in the Hf-Ti system
compared with a previous assessment\cite{1997Bit}]{
        \label{fig:hftih}
        \includegraphics[width=3.0in]{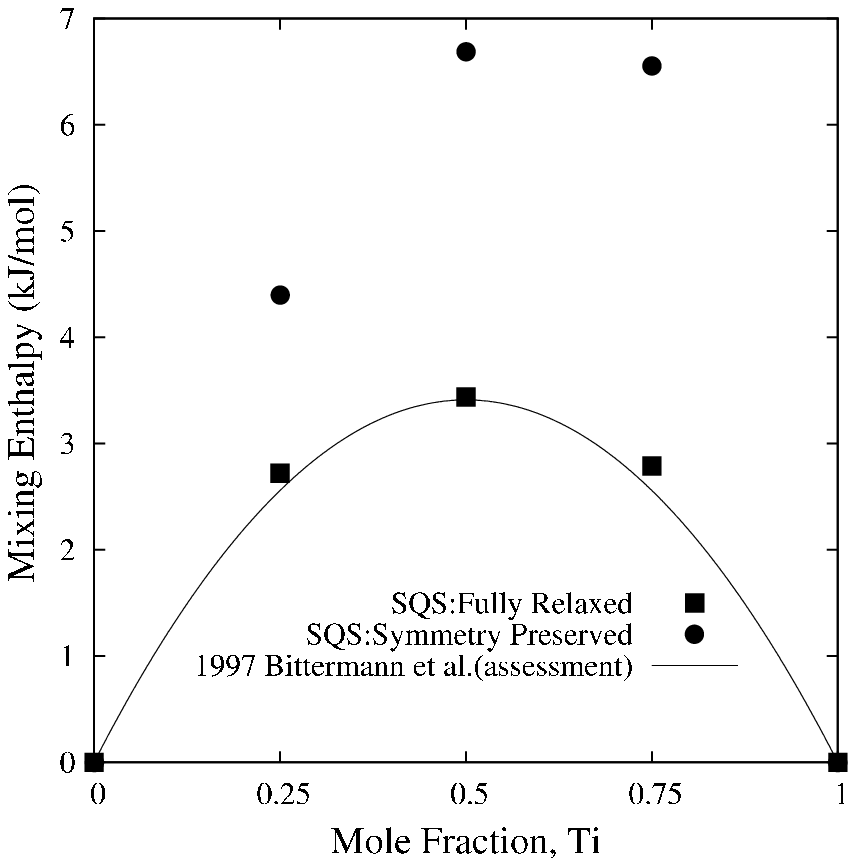}
    }
    \subfigure[~Calculated enthalpy of mixing for the hcp phase in the Ti-Zr system
compared with a previous assessment\cite{1994Kum}]{
        \label{fig:tizrh}
        \includegraphics[width=3.0in]{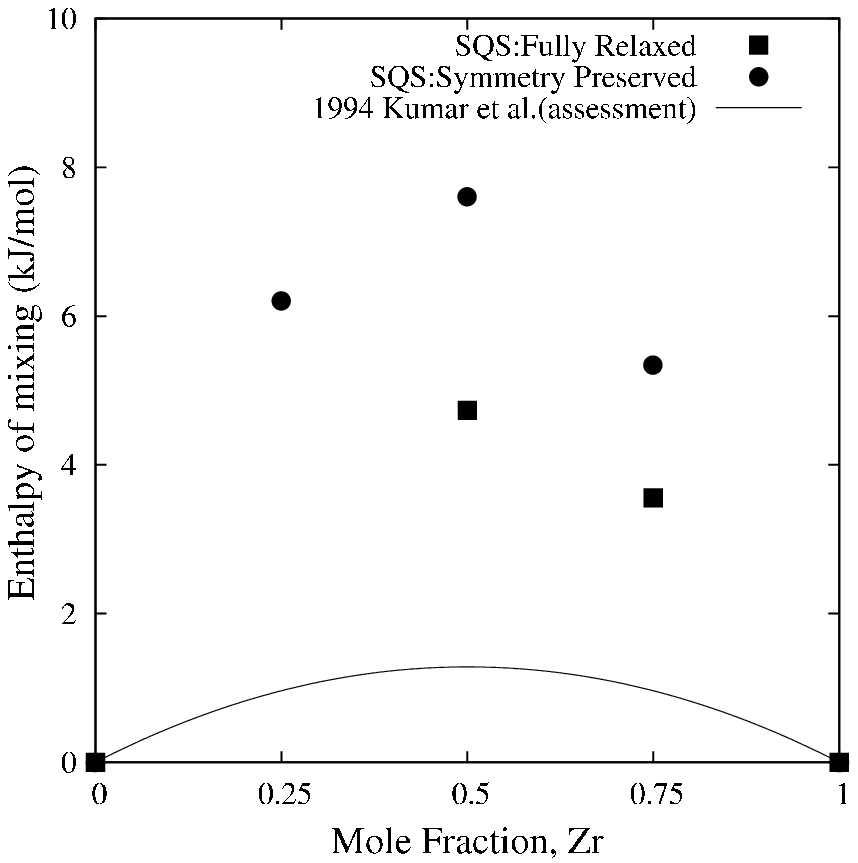}
    }\\
    \subfigure[~Calculated enthalpy of mixing for the hcp phase in the Hf-Zr
system. $\Delta H_{mix} \simeq 0$]{
        \label{fig:hfzrh}
        \includegraphics[width=3.0in]{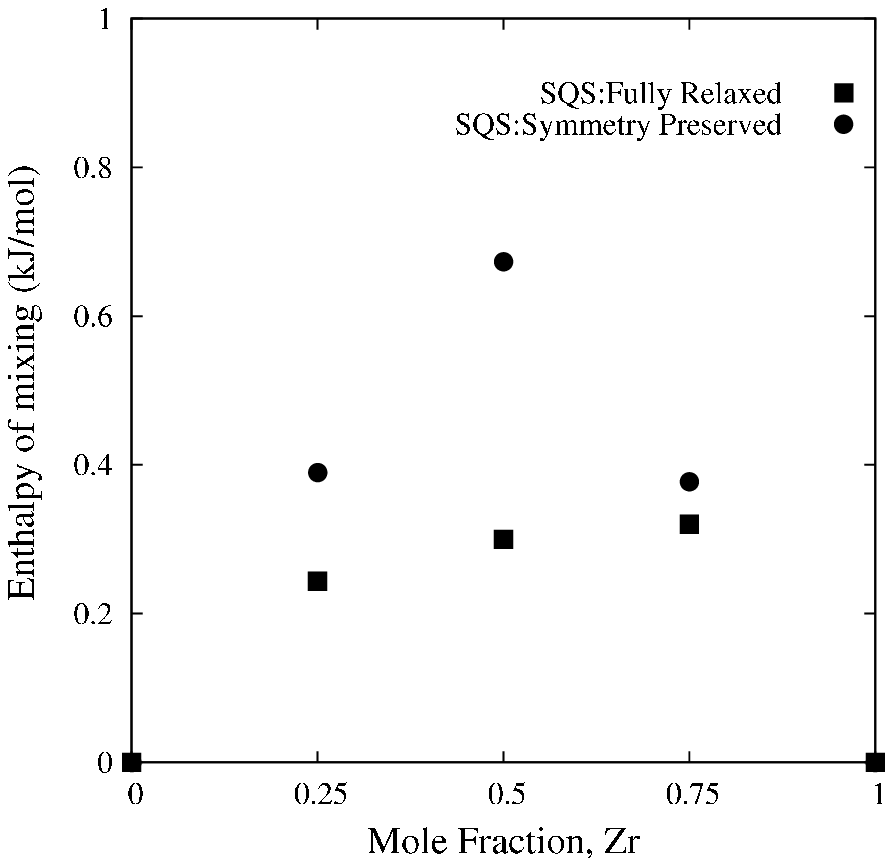}
    }
\caption{Enthalpy of mixing for the Hf-Ti, Hf-Zr and Ti-Zr  binary hcp
solutions calculated from first-principles calculations and CALPHAD
thermodynamic models. All the reference states are hcp structures.}
\end{figure*}

The group IVA transition metals, Ti, Zr, and Hf have hcp structure at low
temperatures and transform to bcc at higher temperatures due to the effects of
anharmonic vibrations. When they form a binary system with each other, they
show complete solubility for both the hcp and bcc solutions without forming any
intermetallic compound phases in the middle.

The Hf-Ti binary is reported to have a low temperature miscibility gap and was
modeled with a positive enthalpy of mixing by \citet{1997Bit}. \fig{fig:hftih}
shows remarkable agreement between the fully relaxed first-principles
calculations and the thermodynamic model, which was obtained by fitting the
experimental phase boundary data. Despite the fact that the local relaxation
energies are rather large ($\sim4$ kJ/mol), the lattice parameters in both
cases agree between each other and with the experimental
results\cite{1959Tyl,1966Cha,1969Rud}.

In the case of the Ti-Zr binary, although no low-temperature miscibility gap
has been reported, \citet{1994Kum} found that the enthalpy of mixing for the
hcp solutions in this binary was positive through fitting of phase diagram
data. Our results confirm this finding, although with even more positive
enthalpy. They are in fact similar in value to those calculated in the Hf-Ti
alloys, suggesting that a low temperature miscibility gap may also be present
in this binary.

In the Hf-Zr system no miscibility gap has been reported. The hcp phase was
modeled as an ideal solution ($\Delta H_{mix}=0$) in the CALPHAD
assessment\cite{2002Bit}. The present calculations suggest that the enthalpy of
mixing of this system is positive, although rather small. In this case, it is
expected that any miscibility gap would only occur at very low temperatures.

The three systems described in this section are chemically very similar, having
the same number of electrons in the \emph{d} bands. Electronic effects due to
changes in the widths and shapes of the DOS of the \emph{d} bands are not
expected to be significant in determining the alloying energetics. Charge
transfer effects are also expected to be negligible. The enthalpy observed can
then be explained by just considering the atomic size mismatch between the
different elements. As was shown in \tab{tbl:bond}, the Hf-Zr hcp alloys are
the ones with the smallest difference in their lattice parameter, thus
explaining their very small positive enthalpy of mixing.

\begin{figure*}[htb]
\centering
    \includegraphics[width=5in]{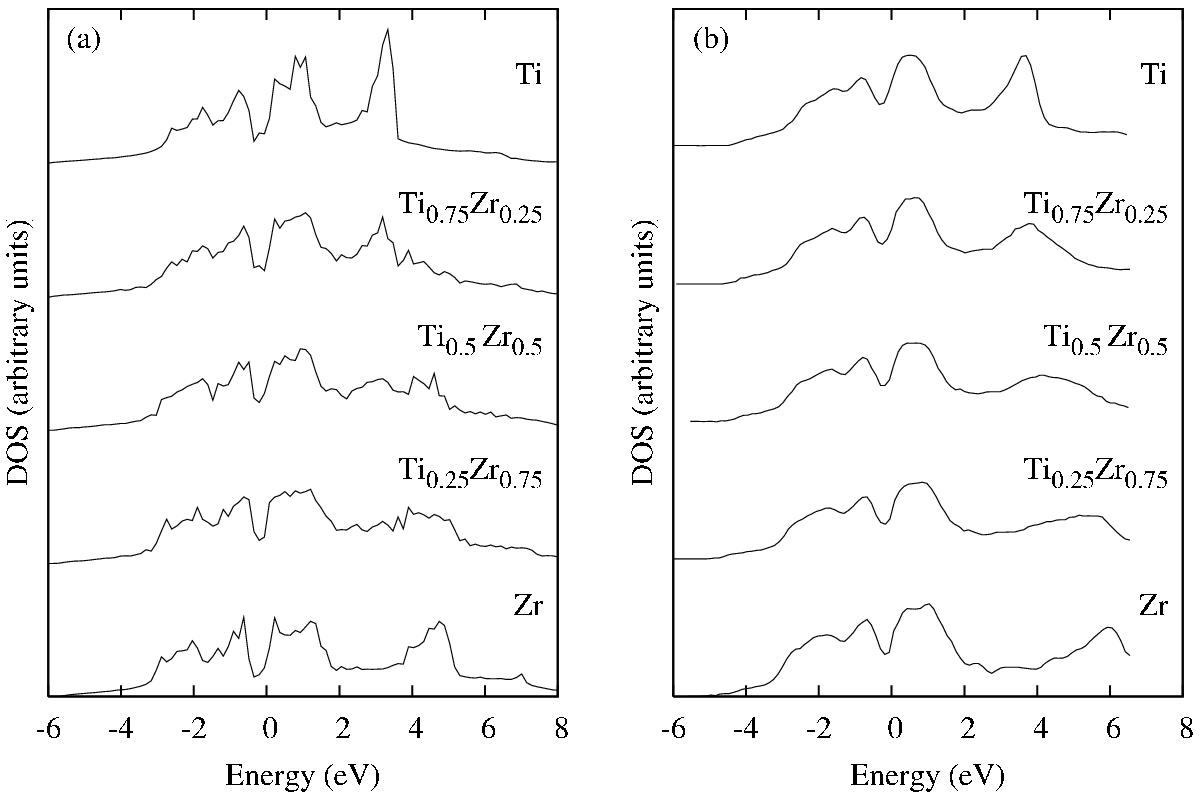}
\caption{Calculated DOS of $\rm Ti_{1-x}Zr_x$ hcp solid solutions from (a) SQS
and  (b) CPA\cite{1991Kud}} \label{fig:tizr_dos}
\end{figure*}

As a final analysis of the ability of the generated SQS to reproduce the
properties of random hcp alloys, \fig{fig:tizr_dos} shows the alloying effects
on the electronic DOS in Ti-Zr hcp alloys. The figure also presents the results
obtained through the CPA approach by \citet{1991Kud}. As can be seen in the
figure, both calculations predict that the DOS corresponding to the occupied d
states are virtually insensitive to alloying. The overall shape of the
\emph{d}-DOS remains relatively invariant. Since Ti and Zr have the same number
of valence electrons, the fermi level remains essentially unchanged as the
concentration varies from pure Zr to pure Ti. On the other hand, alloying
effects are more pronounced in the \emph{d}-DOS corresponding to the unoccupied
states. \fig{fig:tizr_dos} shows how the broad peak at $\sim4.5\,eV$ of the
\emph{d}-DOS for Zr is gradually transformed into a narrow peak at
$\sim3.0\,eV$ as the Ti content in the alloy is increased. The results from the
CPA and the first-principles SQS calculations thus agree with each other,
confirming the present results.

\section{Summary}\label{sec:summary}

We have created periodic special quasirandom structures with 16 atoms for
binary hcp substitutional alloys at three different compositions, 25, 50, and
75 at.\%, to mimic the pair and multi-site correlations of random solutions.

The generated SQS's were tested in seven different binaries and showed fairly
good agreement with existing experimental either enthalpy of mixing and/or
CALPHAD assessments and lattice parameters. Analysis of the radial distribution
and bond lengths in the 21 calculated SQS's, yielded a detailed account of the
local relaxations in the hcp solutions and has been proven a useful way of
characterizing the degree relaxation over several coordination shells.

It should also be noted that when using enthalpy of mixing to derive formation
enthalpy to compare with experimental measurements, there can be a severe
discrepancy between theoretical calculations and experimental data when the
lattice stability, or structural energy difference, from first-principles
calculation is problematic such as the Mo-Ru system in this work. This problem
remains as an unsolved issue.

These supercells can be applied directly to any substitutional binary alloys to
investigate the mixing behavior of random hcp solutions via first-principles
calculations without creating new potentials, as in the coherent potential
approximation (CPA) or calculating other structures in the cluster expansion.
Although the size of the current SQS's is not large enough to generate a
supercell which can satisfy its correlation function at more than just three
compositions ($x$ = 0.25, 0.5, and 0.75 in $A_{1-x}B_x$ binary), calculations
for these compositions can yield valuable information about the overall
behavior of the alloys.

\section*{Acknowledgements}
The authors acknowledge financial support from National Science Foundation
(NSF). First-principles calculations were carried out on the LION clusters at
the Pennsylvania State University. The authors would like to thank Dr.
Christopher Wolverton at Ford for critical proof reading of the manuscript. Dr.
Earle Ryba is acknowledged for his valuable advice for the radial distribution
analysis.

\end{document}